\documentclass[ twocolumn,
aps,prd,   
               preprintnumbers,numbers,sort&compress,nofootinbib,
                            showpacs,
               colorlinks,
               linkcolor=blue,   
               citecolor=blue]{revtex4-1}
\newcommand{\exclude}[1]{}
\usepackage{graphicx}% Include figure files
\usepackage{dcolumn}% Align table columns on decimal point
\usepackage{bm}% bold math
\usepackage{hyperref}

\usepackage{amsmath}
\usepackage[caption=false]{subfig}

\def\ra{\rangle}
\def\la{\langle}
\newcommand{\be}{\begin{eqnarray}}
\newcommand{\ee}{\end{eqnarray}}
\begin{document}
\title{The Axion Quark Nugget Dark Matter Model:  Size Distribution and Survival Pattern  }

\author{Shuailiang Ge}
  \email{slge@phas.ubc.ca}
  \author{Kyle Lawson}
  \email{klawson@phas.ubc.ca}
 \author{Ariel  Zhitnitsky}
  \email{arz@phas.ubc.ca}
  
 \affiliation{Department of Physics \&  Astronomy, University of British Columbia, 
Vancouver,  Canada}

\begin{abstract}
We consider the formation and evolution of Axion Quark Nugget dark matter particles in the 
early universe. The goal of this work is to estimate the mass distribution of these objects 
and assess their ability to form   and survive  to the present day. We argue that this model  
allows a broad range of parameter space in which the AQN may account for the observed dark 
matter mass density, naturally  explains a similarity between the ``dark" and ``visible" components, i.e. $\Omega_{\rm dark}\sim \Omega_{\rm visible}$,  and also offer an  explanation for a number of 
other long standing puzzles such as ``Primordial Lithium Puzzle" and ``the Solar Corona Mystery" 
among many other cosmological puzzles. 
\end{abstract}

\maketitle

\section{Introduction}
In this paper we describe a scenario in which the dark matter consists of macroscopically large, 
nuclear density, composite objects known as  axion quark nuggets (AQN) \cite{Zhitnitsky:2002qa}. 
In this model the ``nuggets" are composed of large numbers of standard model quarks bound 
in a non-hadronic high density colour superconducting (CS) phase. As with other high mass dark matter candidates (such as 
Witten's quark nuggets \cite{Witten:1984rs}, see \cite{Madsen:1998uh} for review) these objects are ``cosmologically dark" not through the weakness of their 
interactions but due to their small cross-section to mass ratio which scales all observable consequences. 
As such, constraints on this type of dark matter place a lower bound on their mass 
distribution, rather than coupling constant.  

There are two additional elements in AQN model in comparison with the older well-known construction \cite{Witten:1984rs,Madsen:1998uh}. 
First, there is an additional stabilization factor  provided by the axion domain walls
which are copiously produced during the QCD transition and which help alleviate a number of 
the problems inherent in the older models\footnote{\label{first-order}In particular, the first order phase transition was a required feature of the system for the original nuggets  to be formed during the QCD phase transition.  However it is known by now that the QCD transition is a crossover rather than the first order phase transition. Furthermore, the nuggets  \cite{Witten:1984rs,Madsen:1998uh}
will likely evaporate on the Hubble time-scale even if they had been formed. In  case of the AQNs the first order phase transition is not required  as the axion domain wall plays the role of the squeezer. Furthermore, the argument  related to the fast evaporation of the nuggets    is not applicable for the  AQNs  because the  vacuum ground state energies inside (CS phase) and outside (hadronic phase)  the nuggets are drastically different. Therefore these two systems can coexist only in the presence of the additional external pressure provided by the axion domain wall, in contrast with original models \cite{Witten:1984rs,Madsen:1998uh} which  must  be stable at zero external pressure.}.  Another  
crucial    additional  element in the proposal      is that the nuggets could be 
made of matter as well as {\it antimatter} in this framework. This novel  key element of the model \cite{Zhitnitsky:2002qa} completely changes entire framework 
because the dark matter density  $\Omega_{\rm dark}$ and the baryonic matter density $ \Omega_{\rm visible}$ now become intimately related to each other and proportional to each other  $\Omega_{\rm dark}\sim \Omega_{\rm visible}$ irrespectively of any specific details of the model, such as the axion mass or size of the nuggets. Precisely this fundamental consequence of the model was the main  motivation for its construction.

The presence of a large amount of antimatter in the form of high density AQNs leads to a large  
number of observable consequences  within  of this model as a result of  annihilation events between antiquarks from AQNs and visible baryons. We refer to next section \ref{AQN} for a short overview 
of the  basic  results,  accomplishments and constraints of this model. 

The only comment we would like to make here  is that 
some  long standing problems may find their natural resolutions within AQN framework.
The first of these is the ``Primordial Lithium Puzzle" which has persisted for at least two 
decades.  It has been recently shown \cite{Flambaum:2018ohm} that  
this long standing mystery  might be  naturally resolved within the AQN scenario. 
Another example  is the 70 year old mystery (since 1939) known in the community under 
the name ``the Solar Corona Mystery". 
It has been recently suggested that this mystery may also find its natural resolution within the 
AQN scenario \cite{Zhitnitsky:2017rop,Raza:2018gpb} as a result of the annihilation of AQNs 
in the solar corona. These two examples show   a   very broad  application potential  of this model.  Furthermore, the corresponding quantitative results are highly sensitive to the size distribution of the AQNs, and their ability to survive in unfriendly environment such as solar corona or high temperature plasma during the big bang nuclear-synthesis (BBN). 

The main goal of the present work is to focus on these two specific questions of the model which have been previously ignored, mostly due to the oversimplified settings with the main goal of  qualitative 
(order of magnitude estimate) rather than quantitative  description.
We are now in a position to fill this gap  and address the hard questions on size distribution 
and survival pattern during the long evolution of the Universe.

% In this work we will consider the formation history of AQN dark matter and estimate 
%the resulting mass distribution. 
The central result being a demonstration that AQN of a sufficient size 
will survive the high density plasma of the early universe from their formation at the QCD 
transition until the present day. However, before turning to the details of formation and 
evolution we will briefly review the relevant properties of the AQN dark matter model, its basic predictions, results and accomplishments  in Section \ref{AQN}.  In sections \ref{sec:RTE} and \ref{sec:distribution} we discuss the size distribution during the formation period, while sections \ref{sec:survival},
 \ref{ssec:PreBBN}, \ref{low-T} and \ref{recombination}  are devoted to an analysis of the survival features of the AQNs at high (before the BBN epoch) and low (after the BBN epoch) temperatures. In section  \ref{present}  we analyze the  present day observational constraints on the mass distribution. 

\section{The AQN Dark Matter Model}\label{AQN} 
\subsection{The basic predictions, results and accomplishments}
The AQN dark matter model was originally introduced to resolve two 
important outstanding problems in cosmology: the nature of dark matter and the mechanism of 
baryogenesis.   The connection between these seemingly 
unrelated questions is motivated by the apparently coincidental similarity of the visible and dark 
matter energy densities,
\begin{equation}
\label{eq:vis-dark}
\Omega_{\rm dark}\sim  \Omega_{\rm visible}.
\end{equation}
If the dark matter is in fact a new fundamental particle then there is no a priori reason for this 
similarity and the visible and dark matter could have formed with widely different energy 
densities. If however these two forms of matter share a common origin the ratio (\ref{eq:vis-dark}) 
may have a physical explanation rather than being a tuned parameter of the theory. In the case 
of the visible matter the energy density is fixed at the QCD   transition when baryons 
form and acquire their observed mass\footnote{Prior to the QCD  transition the quarks and 
leptons carry masses generated via the Higgs mechanism, but these are three orders of magnitude
smaller than the baryon masses and thus represent a negligible fraction of the present day energy 
density.} so we may ask if the dark matter density may also form at this time. 

The AQN proposal  represents an 
alternative to baryogenesis  scenario  when  the 
``baryogenesis'' is replaced by  a charge separation process 
in which the global baryon number of the Universe remains 
zero. In this model the unobserved antibaryons  come to comprise 
the dark matter in the form of dense antinuggets in a colour superconducting (CS) phase.  Dense 
nuggets in a CS phase also present in the system
such that the total baryon charge remains zero at all times during the evolution of the Universe. The detail mechanism of the formation of the  nuggets 
and antinuggets has been recently developed in refs.   \cite{Liang:2016tqc,Ge:2017ttc,Ge:2017idw}.  The only comment we would like to make here is that the energy per baryon charge   is approximately the same 
for nuggets in CS phase and the visible matter  in hadronic phase as the both types of matter are formed during the same QCD transition, and both are proportional to the same fundamental dimensional  parameter  $\sim \Lambda_{\rm QCD}$. Therefore, the relation (\ref{eq:vis-dark}) is a natural outcome of the framework rather than a consequence of a fine-tunning.  

\begin{figure}
\centering
\captionsetup{justification=raggedright}
\includegraphics[width=0.8\linewidth]{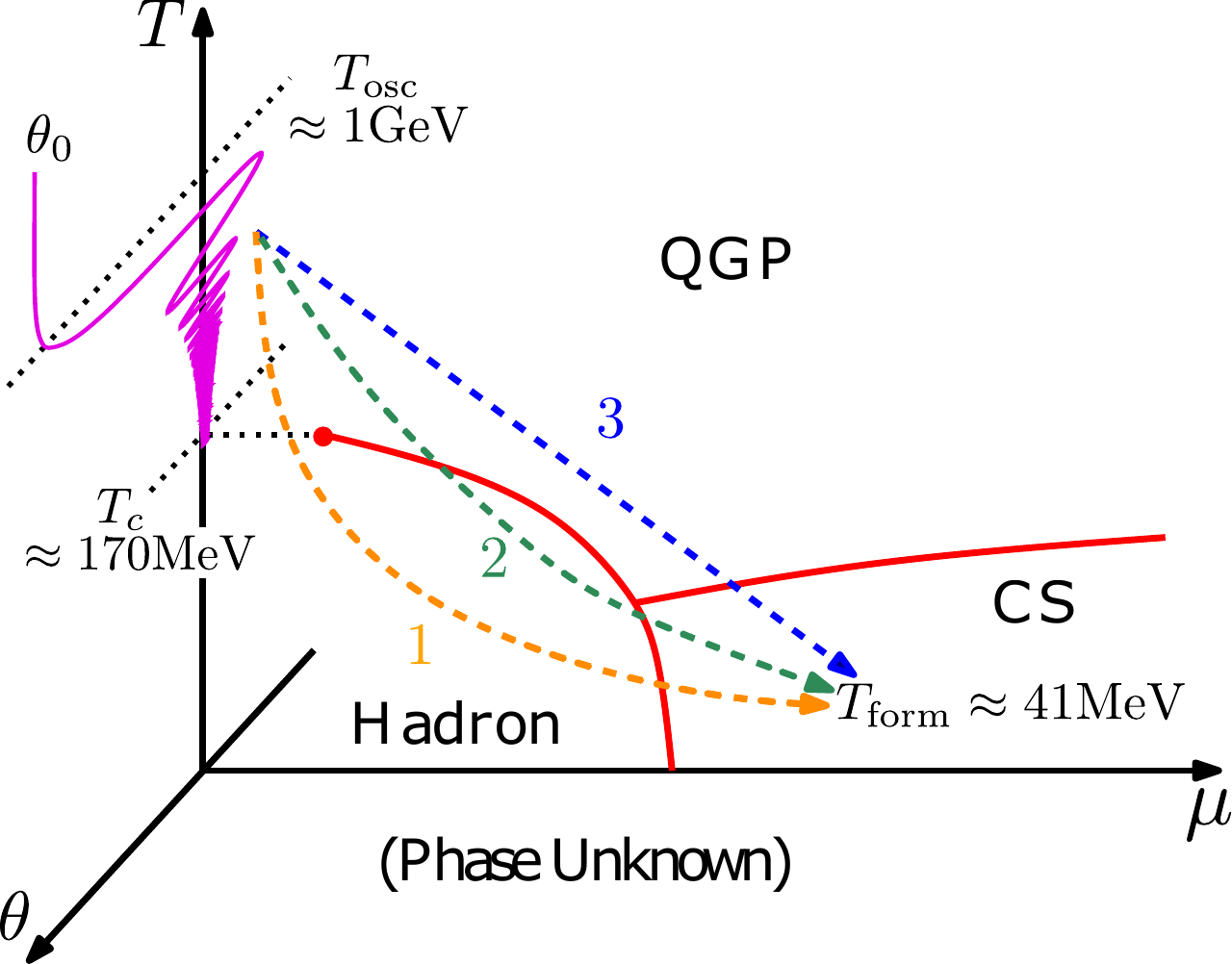}
\caption{This diagram illustrates the interrelation between the axion production due to the misalignment mechanism and the nugget's formation which starts before the axion field $\theta$ relaxes to zero. Adopted from \cite{Ge:2017idw}.}
\label{phase_diagram}
\end{figure}

In the context of the AQN model the dark matter is formed by the action of a collapsing network 
of axion domain walls formed at the QCD  transition. These processes will contribute to the direct axion production through misalignment mechanism,  domain wall decay, as well as the nugget's formation,   see Fig.\ref{phase_diagram}. This process will be described in 
detail in section \ref{sec:RTE}, and we shall not elaborate on this topic here. 

\exclude{The basic picture is that any closed bubble walls will sweep up 
the baryonic matter through which they move concentrating it to higher densities. However, due to 
the known \textsc{cp} violation along the axion walls a closed bubble may compact either matter or 
antimatter more efficiently. In fact the matter-antimatter asymmetry is known to be an order one 
effect in this case. Compression continues until the bubble's collapse is halted 
by the internal fermi pressure and the remaining high density baryons or antibaryons become the 
AQN. 
}

The result of this ``charge separation"  process is two populations of AQN carrying positive and 
negative baryon number. That is the AQN may be formed of either matter or antimatter. 
However, due to the global  $\cal CP$ violating processes associated with $\theta_0\neq 0$ during 
the early formation stage, see   Fig.\ref{phase_diagram},  the number of nuggets and antinuggets 
formed will be different. This difference is always an order of one effect   irrespectively to the 
parameters of the theory, the axion mass $m_a$ or the initial misalignment angle $\theta_0$, as argued in  \cite{Liang:2016tqc,Ge:2017ttc}.  
         The disparity between nuggets $\Omega_{N}$ and antinuggets $\Omega_{\bar{N}}$ unambiguously implies that  the  baryon contribution $\Omega_{B}$ must be the same order of magnitude as all contributions are proportional to one and the same dimensional parameter $\Lambda_{\rm QCD}$. 
         
         If we assume that all other dark matter components (including conventional axion production as shown on  Fig. \ref{phase_diagram}) are subdominant players, one can relate 
         the baryon charge hidden inside the nuggets   with the visible baryon charge.  Indeed, the observations suggest that $\Omega_{\rm dark} 
        $ is 5 times greater than $\Omega_{B}$ which in our framework implies that
\begin{equation}
\label{eq:ratio}
\Omega_{\bar{N}}:\Omega_{N}:\Omega_{B} \approx 3:2:1, ~~  \Omega_{\rm dark} \approx (\Omega_{\bar{N}}+\Omega_{N}).
\end{equation}
This approximate relation  represents a direct consequence  of the baryon charge conservation
\be
  \label{eq:321}
    B_{\text{universe}} = 0 &=& B_{\text{nugget}}
    + B_{\text{visible}}- |{B}|_{\text{antinugget}}. 
     \ee
     If the direct axion production is not negligible the ratio  (\ref{eq:ratio}) will be obviously  modified. However, all numerical coefficients entering (\ref{eq:ratio}) will always be order of one, unless the axion mass $m_a$ is fine tuned  to saturate the  present value for $\Omega_{\rm dark}$. We refer to the original paper  \cite{Ge:2017idw} and specifically Fig. 5 in that paper for a more precise 
     relation between these two distinct contributions.  One should note that  the direct axion production contribution  $\Omega_{\rm axion}$  to dark matter  $\Omega_{\rm dark}$  is highly sensitive to the axion mass $m_a$ in contrast with 
     nugget's contribution (\ref{eq:vis-dark}) which holds irrespectively to the value of the axion mass $m_a$.
   In particular,   $\Omega_{\rm axion}$  becomes negligible  for sufficiently large axion mass because it scales as $\Omega_{\rm axion}\sim m_a^{-7/6}$, see footnote \ref{defects}   for recent numerical estimates.  These estimates suggest that if $m_a\geq  10^{-4} {\rm eV}$ the direct axion production is numerically small $\Omega_{\rm axion} \ll \Omega_{\rm dark}$ and, therefore,   the ratio (\ref{eq:ratio}) is approximately valid.

We may also reformulate   expression (\ref{eq:ratio})  in terms of the number densities and average 
mass of the various components,
\begin{equation}
\frac{1}{3}  \langle M_{\bar{N}} \rangle n_{\bar{N}} \approx 
\frac{1}{2} \langle M_N \rangle n_N \approx m_B n_B
\end{equation}
with the AQNs masses related to their baryon number by $M_N \approx M_{\bar{N}}\approx m_p |B|$. 
The resulting AQN will be macroscopically large (typically with radii above $10^{-5}$cm) and 
of roughly nuclear density resulting in masses above roughly a gram. 
The density of the 
colour superconducting phase is not precisely known and  depends on the exact phase of matter 
realized in the AQN. Furthermore, the axion domain wall surrounding the nugget also contributes to its mass, see   \cite{Ge:2017idw} for quantitative relations. For the present work we will simply adopt a typical  nuclear baryon number density of order $10^{40} {\rm cm}^{-3}$ for all our estimates  
 such that a nugget with $|B| \sim 10^{25}$ has a typical radius 
of $R\sim 10^{-5}$cm. 

As a result,  the effective interaction is very small $\sigma_N/M_{N}\sim10^{-10}{\rm cm}^2/{\rm g}$ 
where  $\sigma_N\sim R^2$ assumes a typical geometrical cross section. This estimate is  
well below the upper limit of the conventional DM constraint $\sigma/M_{\rm DM}<1{\rm cm}^2/{\rm g}$. This is the main reason why despite being made from strongly interacting particles, the AQN 
nevertheless  behave as cold DM  from the cosmological perspective.

 Another fundamental ratio 
is the baryon to entropy ratio at present time
\be
\label{eta}
\eta\equiv\frac{n_B-n_{\bar{B}}}{n_{\gamma}}\simeq \frac{n_B}{n_{\gamma}}\approx 6\times 10^{-10}.
\ee
In  the AQN proposal (in contrast with conventional baryogenesis frameworks) this ratio 
is determined by the formation temperature $T_{\rm form}\simeq 41 $~MeV  at which the nuggets and 
antinuggets complete their formation, see Fig. \ref{phase_diagram}.  We refer to \cite{Liang:2016tqc} for relevant estimates
of the parameter $\eta$ within the AQN framework. We note that 
$T_{\rm form}\sim \Lambda_{\rm QCD}$   assumes a typical QCD value, as it should. This is because  
there are no small parameters in QCD and all observables must be expressed in terms of   a single fundamental parameter which is the $\Lambda_{\rm QCD}$. One should add here that the numerical smallness of the factor (\ref{eta})    is a  result of an exponential sensitivity of $\eta$  to the formation temperature as $\eta\sim \exp (- {m_p}/{T_{\rm form}})$ with   the proton's mass being  numerically  large  parameter when it  is written in terms of the QCD critical temperature $m_p\simeq 5.5 T_c$.  

It is the purpose of 
this work to investigate the mass distribution of AQN generated by the collapse of the axion 
domain wall network and assess their survival pattern within the high density plasma of the early universe. 
The efficiency of the formation process 
%and thus the degree of baryogenesis 
is reflected in the 
observed baryon to photon ratio (\ref{eta}), 
which implies that only a small fraction of the primordial baryonic matter  is successfully bound when the formation is completed at $T\approx 41$ MeV 
into AQN and thus protected from further annihilation.   

%For the purposes of the calculations that 
%follow it should be noted that the value of $\eta$ was slightly higher prior to the annihilation of 
%$e^+e^-$ pairs, though its exact numerical value does not seriously effect the arguments which follow. 
To conclude this short overview on basic features of the AQN framework we would like to mention that 
the AQN dark matter model has recently been applied to a variety of situations in the 
early universe. In particular it has been suggested that the anomalously strong 21 cm absorption  
feature reported by the EDGES collaboration \cite{Bowman:2018yin} may be driven by an additional 
component to the large wavelength end of the radiation spectrum at early times \cite{Feng:2018rje}. 
Such a component may be produced by thermal emission from a population of AQN which primarily 
emit well below the CMB peak and which are not expected to be in thermal equilibrium with the 
photons \cite{Lawson:2018qkc}. 

It has also been suggested that the presence of partially ionized 
AQN at temperatures $T\simeq 20 $ keV soon after  the BBN  epoch  may result in the preferential 
capture (and eventual annihilation) of the  highly charged heavy nuclei with $Z\geq 3$ produced during BBN.   
This proposal  offers a resolution \cite{Flambaum:2018ohm} to the long standing problem coined as the  ``Primordial Lithium Puzzle".   While both these phenomena which occurred at very early times in the evolution of the Universe (at the redshift $z\simeq 17$ and $T\simeq 20 $ keV correspondingly) are potentially very interesting and important,   the underlying physics and astrophysical backgrounds are not sufficiently 
well understood to impose strong constraints on the AQN size distribution during these earlier times. Instead the strongest constraints 
come from present, more readily observable and better understood environments and will be the subject of 
the next subsection \ref{sec:Constraints} of this work.  

\subsection{Mass distribution constraints}\label{sec:Constraints}
As stated above the nuggets are not fundamentally weakly interacting but are  effectively ``dark" 
due to their large mass and consequent low number density. For example the flux of nuggets 
that should be observed on or near earth is,
\begin{equation}
\label{eq:flux}
\Phi = n_N v_N \approx \frac{\rho_{DM} v_N}{M_N} \approx 1~{\rm{km}}^{-2}{\rm{yr}}^{-1}
\left( \frac{10^{24}}{\langle B \rangle} \right)
\end{equation} 
so that direct detection experiments impose lower limits on the value of 
$\langle B \rangle$ for the distribution of the AQN. Limits may also be obtained from 
astrophysical and cosmological observations. In this case any observable consequences 
will be scaled by the matter-AQN interaction rate along a given line of sight,
\begin{equation}
\label{eq:observable}
\Phi \sim R^2\int d\Omega dl [n_{\rm visible}(l)\cdot n_{DM}(l)] \sim \frac{1}{\langle B \rangle^{1/3}},
\end{equation}
where $R\sim B^{1/3}$ is a typical size of the nugget which determines the effective cross section of interaction between DM and visible matter. 
Thus, as with direct detection, astrophysical constraints impose a lower bound on the value of 
$\langle B \rangle$. 

The relevant constraints come from a variety of both direct detection and astrophysical observations 
which we will list briefly here. Further details are available from the original papers  and references  therein.

\subsubsection{Direct Detection}\label{ssec:DD}
As mentioned above the flux of AQN on the earth's surface is scaled by a factor of $B^{-1}$ 
and is thus suppressed for large nuggets. For this reason the experiments most relevant to AQN detection 
are not the conventional high sensitivity dark matter searches but detectors with the largest 
possible search area. For example it has been proposed that large scale cosmic ray detectors 
such as the Auger observatory of Telescope Array may be sensitive to the flux of AQN in an 
interesting mass range, however this sensitivity is strongly limited by the relatively low velocity 
of the AQN $v_{N} \sim 10^{-3}c$ \cite{Lawson:2010uz}. 

The strongest direct detection limit 
is likely set by the IceCube Observatory's non-detection of a non-relativistic magnetic monopole
\cite{Aartsen:2014awd}. While the magnetic monopoles and the AQNs interact with material of the detector
in very different way, in both cases the interaction leads to the electromagnetic and hadronic cascades along the trajectory 
of AQN (or magnetic monopole) which must be observed by the  detector if such event occurs.  A non-observation of any such cascades puts a limit on the flux of heavy non-relativistic particles passing through the detector,  
which limits the AQN flux to $\Phi_N \lesssim 1$km$^{-2}$yr$^{-1}$ which is mainly determined by the size of the IceCube Observatory.  Similar limits are also 
derivable from the Antarctic Impulsive Transient Antenna  (ANITA) \cite{Gorham:2012hy} though 
this result   depends  on the details of radio band emissivity of the AQN. There is also a 
constraint on the flux of heavy dark matter with mass $M<55$g based on the non-detection of 
etching tracks in ancient mica \cite{Jacobs:2014yca}. 

If we take the local dark matter mass density to be $\rho \approx 0.3$ GeV/cm$^{3}$ and assume 
that this value is saturated by the AQN we may translate the flux constraint $\Phi_N \lesssim 1$km$^{-2}$yr$^{-1}$ 
into a lower limit with $3.5\sigma$ confidence on the mean baryon number of the nugget distribution of 
\be
\label{direct}
\la B \ra > 3\cdot 10^{24} ~~~~~({\rm direct ~ detection ~constraint)}, 
\ee
where we assume   100 \% efficiency of the observation of the AQNs passing through IceCube Observatory.
If this efficiency is   lower the  limit (\ref{direct}) will be also weakened correspondingly.  

\subsubsection{Indirect Detection}\label{ssec:ID}
We also consider the constraints arising from possible dark matter interactions within the 
solar system. These include a limit from potential contribution to earth's energy budget which 
require $|B| > 2.6\times 10^{24}$ \cite{Gorham:2012hy}, which is consistent with (\ref{direct}). It has also been suggested that there 
is a strong limit on any annihilating AQN population due to the low flux of high energy neutrinos 
from the sun \cite{Gorham:2015rfa}. However the composite nature of the AQN when the bulk of quark matter 
is in CS phase is likely to result 
in the majority of neutrino emission to occur at relatively low energies where they will be lost in the 
much larger conventional solar neutrino background \cite{Lawson:2015cla}.

\subsubsection{Galactic Observations}\label{ssec:Gal}
It is known  that the  spectrum from galactic center (where the dark and visible matter densities assume the high   values)
contains several excesses of diffuse emission the origin of which 
is unknown, the best 
known example being the strong galactic 511~KeV line  \cite{Prantzos:2010wi}.

The emission spectrum of the AQN has been studied in a variety of environments and at a 
range of energy scales. As discussed above the expected diffuse emission due to the AQN 
scales as the product of the dark and visible matter densities ($\rho_{DM}\cdot \rho_B$) so 
that the strongest consequences will be from relatively high density regions such as the galactic 
centre. The dependance on the visible matter density also means that the AQN model predicts 
a dark matter contribution to the galactic spectrum that is less spherically symmetric than is 
expected for either decaying ($\sim \rho_{DM}$) or self annihilating ($\sim \rho_{DM}^2$) dark 
matter models, consistent with observations   \cite{Prantzos:2010wi}. The emission spectrum associated with an AQN population will be distinct from 
that of more conventional dark matter candidates in that the most energetic annihilations will be at 
the nuclear ($\sim 100$ MeV) scale implying that any signal will be limited to  sub-GeV energies. 

The potential AQN contribution to the galactic spectrum has been analyzed across a 
broad range of frequencies from a radio band thermal contribution to 
x-ray and $\gamma$-ray photons produced by more energetic annihilation events. 
In each case the predicted emission was consistent 
with observations and had the possibility of improving the global fit of galactic emission models. 
All these  emissions from different frequency bands  are expressed in terms of the same integral (\ref{eq:observable}), and therefore, the  relative  intensities  are unambiguously and completely determined by internal structure of the nuggets which is described by conventional nuclear physics and basic QED. For further details see the original works 
\cite{Oaknin:2004mn, Zhitnitsky:2006tu,Forbes:2006ba, Lawson:2007kp,
Forbes:2008uf,Forbes:2009wg,Lawson:2012zu}   with specific computations in different frequency bands in galactic radiation, and a short overview
\cite{Lawson:2013bya}.  

To summarize this subsection: the most 
significant potential dark matter signal in this context is the galactic 511 keV line  (plus related continuum emission in $100$ keV range)   resulting from 
low energy electron-positron annihilations through the $^1S_0$ and $^3S_1$ positronium formation with consequent decays. These emission features of the galactic spectrum have 
proven difficult to explain with conventional astrophysical sources. At the same time, the AQN model with the constraints from direct detection discussed in section \ref{ssec:DD} 
could offer a potential explanation for  the entire observed 511 keV emission feature  (including the width, morphology, $3\gamma$ continuum spectrum, etc).
\exclude{
 and requires a positron 
source capable of producing annihilations at a rate of $10^{43}$s$^{-1}$ within the galactic 
bulge \cite{Prantzos:2010wi}. Following the arguments of \cite{Oaknin:2004mn, Zhitnitsky:2006tu} 
we can convert this signal into a constraint on the mean AQN baryon number. Beginning with 
expression \ref{eq:observable} and assuming that the dark matter has a central constant density 
core we may write the expected annihilation rate as,
\begin{equation}
\Gamma \approx \frac{\rho_{DM}}{M_N} 4\pi R_N^2 v_g \frac{M_B}{m_p} 
\end{equation}
where $v_g \approx 10^{-3}c$ is the typical galactic velocity and $M_B \approx 10^{10}M_{\odot}$ 
is the visible matter mass contained in the bulge. Assuming that the majority of electrons colliding 
with an antimatter AQN annihilate this gives us a total rate of,
\begin{equation}
\Gamma \approx 10^{43} \frac{\rho_{DM}}{10{\rm{GeV}}\cdot {\rm{cm}}^{-3}} 
\left(\frac{10^{25}}{\la |B| \ra}\right)^{1/3}
\end{equation}
While this result is rather strongly dependent on the poorly constrained distribution of dark matter 
within the galactic centre it emphasizes that an AQN population consistent with the constraints from 
direct detection discussed in section \ref{ssec:DD} could provide the entire observed 511 keV 
emission features. 
}

If further contributions from conventional astrophysical sources are discovered 
the constraint (\ref{direct}) from section \ref{ssec:DD} may be tightened to higher values of $\la B \ra$. As the line of sight through the 
galactic centre samples the emission from a large number of individual AQN this measurement is 
sensitive only to the average baryon number $\la B \ra$ of the antimatter AQN and does not provide any information 
on their size distribution, in contrast with the solar observations discussed in next subsection,  which are highly sensitive to the size distribution. 

\exclude{
\subsubsection{Cosmological Observations}\label{ssec:Cosmo}
In addition to the possible increase in diffuse emission associated with nearby structure it 
is also possible that the AQN may contribute to the diffuse cosmic background due to the 
high densities at early times. We may therefore ask what if there will be an isotropic cosmological 
background associated with the presence of AQN shortly after recombination when the universe 
becomes transparent to low energy radiation. The contribution to the isotropic background of the 
AQN during this epoch was assessed in \cite{Lawson:2012zu} and in \cite{Lawson:2018qkc} in the 
context of the unexpectedly strong 21cm absorption feature observed by EDGES \cite{Bowman:2018yin}. 
The results of these analysis are strongly dependant on the details of thermal emission from the AQN 
}

\subsubsection{Solar Corona Observations}\label{ssec:Solar}
Yet another AQN-related effect might be intimately linked to the so-called ``solar corona heating mystery".
The  renowned  (since 1939)  puzzle  is  that the corona has a temperature  
$T\simeq 10^6$K which is 100 times hotter than the surface temperature of the Sun, and 
conventional astrophysical sources fail to explain the extreme UV (EUV) and soft x ray radiation 
from the corona 2000 km  above the photosphere. Our comment here is that this puzzle  might  find its  
natural resolution within the AQN framework as recently argued in 
 \cite{Zhitnitsky:2017rop,Zhitnitsky:2018mav,Raza:2018gpb}.

In this scenario the AQN composed 
of antiquarks fully annihilate within the so-called transition region (TR) providing a total annihilation energy of order
\begin{equation}
\label{corona}
\Delta E \approx  4\pi b_{\odot}^2 \rho_{\rm DM} v_{\rm DM} 
\approx 5\cdot 10^{27} {\rm{erg/s}},
\end{equation}
where $b_{\odot}$ is the sun's gravitational capture  parameter
 \be
  \label{capture}
  b_{\odot}\simeq R_{\odot}\sqrt{1+\gamma_{\odot}}, ~~~~ \gamma_{\odot}\equiv \frac{2GM_{\odot}}{R_{\odot}v^2}.
  \ee
 The estimate (\ref{corona})  is determined by the 
local dark matter density independent of mass distribution. This value is suggestively close to the 
observed   EUV luminosity of $10^{27}$erg/s. The EUV emission is believed to be powered 
by impulsive heating events known as nanoflares the origin of which is unknown. If these nanoflares 
are in fact AQN annihilating events (which is precisely the main conjecture formulated in \cite{Zhitnitsky:2017rop}
and formally expressed by eq. (\ref{distribution}), see below)  we may extract an upper limit on the mass distribution for the AQNs because the energy distribution of the nanoflares has been previously studied by the solar physics people.  

The main reason for our ability to study  the AQN mass distribution  is due to the fact that the nanoflare distribution has been modelled using magnetic hydro dynamics  (MHD) simulations by plasma and solar physics people to match the solar observations with simulations. We can use the corresponding results to constrain the AQN mass distribution. 

Few comments on nanoflare and AQN distribution are in order. First of all, 
the majority of nanoflares (and therefore,  individual AQN annihilation events) must be below the resolution 
of solar telescopes and they must interact sufficiently with the corona to deposit the majority of 
the available energy in the TR rather than at lower radii. According to \cite{mitra2001nanoflare} 
the resolution limit for flares is $E_{\rm res} \approx 3\times 10^{24}$ erg $\approx 2\times10^{27}m_pc^2$ 
which implies that the majority of AQN must have a baryon number below $B\sim10^{27}$ or, 
alternatively, that only a small fraction of their mass is able to annihilate in the Corona, 
though the later possibility is disfavoured by the analysis of \cite{Raza:2018gpb}. 

Secondly, various analysis of coronal heating based on MHD have considered the nanoflares to be a low energy continuation 
of the higher energy class of solar flares despite the fact that they have a significantly different 
spacial and temporal distributions\footnote{\label{nanoflare}Nanoflares appear to occur uniformly across the solar surface while 
larger flares are strongly correlated with active regions. Furthermore, the EUV intensity (which represents the nanoflare activity) shows very modest variation within the solar cycle. It is in drastic contrast with large flares which demonstrate  a huge variation (factor of $10^2$ or more)  in    frequency of appearance within the solar cycle, see detail discussions  in \cite{Raza:2018gpb}.}. Under this assumption a number of 
energy distributions have been considered generally with power-law fits consistent with the 
better observed population of flares at larger energies. The analysis of \cite{Pauluhn:2006ut} favours 
a power-law with slope $\alpha \approx 2.5$ 
   where $\alpha$ is defined as follows 
  \be
  \label{distribution}
   {dN} \sim  E^{-\alpha} dE \sim B^{-\alpha}dB, 
  \ee
  where $dN$ is the number of the nanoflare  events per unit time with energy between $E$ and $E+dE$.  According to conjecture formulated in \cite{Zhitnitsky:2017rop} this distribution coincides with the baryon charge distribution $dN/dB$ which is the topic of the present work.    These two distributions are tightly linked  
    as these two entities are related to the same AQN objects     according to our  interpretation of the observed nanoflare events.  
 
Third, 
%extending across the nanoflare range and disfavours 
%energy slope with $\alpha>-2$. 
 any population with a slope shallower 
than $\alpha=2$ will experience too few nanoflares to dominate the total heating budget.
However, an alternate analysis \cite{Bingert:2013} considers a broken power-law in which a shallow 
$\alpha \approx 1.2$ slope transitions to a steeper $\alpha\simeq 2.5$ slope at large energies. In this 
case the heating contribution will be peaked at the energy where the break in the spectrum 
occurs, the position of the knee. In the model  \cite{Bingert:2013} it occurs at $E\simeq 10^{24}$ erg,
which is slightly below the instrumental resolution energy $E_{\rm res} \approx 3\cdot 10^{24}$ erg. 
In many respects  we consider this model is preferable from AQN perspective because it explicitly shows 
that nanoflares and flares have different nature, in agreement with indirect evidence pointing to  their distinct origins, see footnote \ref{nanoflare}.

While the mechanism of energy release in the AQN model is substantially different from 
%that in what follows considered in 
conventional flare studies 
%and the AQN need not display the same energy scaling  as the higher energy flare population 
the nanoflare models of \cite{Pauluhn:2006ut} and \cite{Bingert:2013} 
provide a useful parameterization of the distribution of AQN masses which may be consistent 
with the observed degree of coronal heating and EUV emission. 

With this set of observational constraints in mind we now turn to the main purpose of this 
work: a study of the formation and subsequent evolution of the AQN population.

\section{Formation of the AQNs}\label{sec:RTE}
This section should be viewed as a  introduction to domain wall formation mechanism, its basic ideas (such as percolation and formation of the closed surfaces), basic generic results, and main  assumptions. We also overview some results from our previous studies ~\cite{Liang:2016tqc,Ge:2017ttc} which represent the starting point of the quantitative approach which is the subject of  the present work.  We also report some new numerical results (supporting the entire framework)  at the very end of this section.

%The starting point of the nuggets' evolution is the formation of the closed axion domain walls. 
It is known 
that axion domain walls can form in the early Universe~\cite{sikivie1982axions,vilenkin1982cosmic}. 
When the Universe cools down to $T_{\rm osc}\sim 1$~GeV, the axion mass effectively turns on and the 
axion potential gets tilted and the axion field starts to oscilate, see Fig. \ref{phase_diagram}. The tilt becomes much more pronounced at the QCD transition 
$T_{c}\sim 170$~MeV when the chiral condensate forms.  In general, one should expect that the 
axion domain walls can form anywhere between $T_{\rm osc}$ and $T_{c}$, see Fig. \ref{phase_diagram}.

Precisely during this time when $T_{c}<T< T_{\rm osc}$ the axions get  emitted and may contribute to the dark matter density. The conventional mechanism of emission is the  misalignment mechanism \cite{Preskill:1982cy,Abbott:1982af,Dine:1982ah}. The axions may also be radiated   due to the decay of the topological defects \cite{Chang:1998tb,Hiramatsu:2012gg,Kawasaki:2014sqa,Fleury:2015aca,Klaer:2017ond,Gorghetto:2018myk}. In both cases the corresponding contribution is highly sensitive to the axion mass $m_a$ as the corresponding contribution to the dark matter scales as $\Omega_{\rm axion}\sim m_a^{-7/6}$. 
 There is a number of uncertainties and  remaining discrepancies in the corresponding estimates. We shall not comment on these subtleties by referring to the original   papers\footnote{\label{defects}  
  According to the most recent computations presented in ref.\cite{Klaer:2017ond}, the axion contribution to $\Omega_{\rm DM}$ as a result of decay of the  topological objects can saturate the observed DM density today if the axion mass is  in the range $m_a=(2.62\pm0.34)10^{-5} {\rm eV}$, while the earlier estimates suggest that the saturation occurs at a larger axion mass. It could be some other complications with conventional computations as argued in \cite{Gorghetto:2018myk}. One should also emphasize that the computations  \cite{Chang:1998tb,Hiramatsu:2012gg,Kawasaki:2014sqa,Fleury:2015aca,Klaer:2017ond,Gorghetto:2018myk} have been performed with assumption that PQ symmetry was broken after inflation.}.
 
 The  formation of the AQNs always accompanies these two distinct contributions to $\Omega_{\rm DM}$. However, in comparison with the  misalignment mechanism \cite{Preskill:1982cy,Abbott:1982af,Dine:1982ah} and the decay of the topological defects \cite{Chang:1998tb,Hiramatsu:2012gg,Kawasaki:2014sqa,Fleury:2015aca,Klaer:2017ond}
 the contribution of the nuggets to the dark matter is always order of one effect not sensitive  to the axion mass $m_a$ nor to the misalignment angle $\theta_0$ as overviewed in Introduction and expressed by eq. (\ref{eq:vis-dark}).
 The axion field plays a dual role in this framework: it is responsible for the direct production of the propagating axions
 with contribution which scales as $\Omega_{\rm axion}\sim m_a^{-7/6}$.  
 It   also plays a key role in the AQN's formation as discussed in ~\cite{Liang:2016tqc,Ge:2017ttc}.

In the AQN model, we assume the pre-inflation scenario in which the Pecci-Quinn (PQ) phase 
transition occurs before inflation~\cite{Liang:2016tqc}. Normally, in this case no topological 
defects can be formed   as there is a single vacuum state which occupies entire   observable Universe, see footnote \ref{defects} for clarification. 
This argument is absolutely correct for $N_{\rm DW}\neq1$ axion domain walls which require the 
presence of different physical vacua with the same energy. However, $N_{\rm DW}=1$ axion 
domain walls are special in the sense that the axion field $\theta$ interpolates between one and 
the same physical vacuum but corresponding to different topological $k$ branches with 
$\theta\rightarrow\theta+2\pi k$. As explained in the Ref.~\cite{Liang:2016tqc}, different $k$ 
branches of the same vacuum must be present at each point in space to provide the $2\pi$ periodicity 
of the vacuum energy~\cite{Witten:1980sp,Witten:1998uka}. The inflation cannot separate these 
$k$ branches. As a consequence, $N_{\rm DW}=1$ axion domain walls can form even in this 
pre-inflation scenario with $\theta$ interpolating between $k=0$ branch ($\theta=0$) and $k=1$ 
branch ($\theta=2\pi$), see ~\cite{Liang:2016tqc} for the details. 

The key point is that a finite  portion (few percent) of $N_{\rm DW}=1$ walls are formed as closed surfaces. Such behaviour  has been observed in numerous numerical simulations  ~\cite{vachaspati1984formation,Chang:1998tb},  see also 
section~\ref{sec:distribution} for comments and more details. In previous studies  this  contribution to $\Omega_{\rm axion}$ (due to the closed axion domain walls) has been ignored because these closed surfaces (representing only few percent of the total area) collapse 
as a result of  the wall tension and do not play any significant role in the dynamics of the system. However, 
in  the AQN framework this small, but finite portion of the closed surfaces plays a key role.
This is because 
  the collapse of the closed $N_{\rm DW}=1$ bubbles will be halted due to the 
Fermi pressure acted by the accumulated fermions~\cite{Liang:2016tqc}. As a result,  the closed 
$N_{\rm DW}=1$ bubbles will eventually become the stable nuggets and serve as the 
dark matter candidates.

The dynamics of the  nuggets with  size $R(t)$ is governed  by the following equation \cite{Liang:2016tqc}, 
\begin{equation}
\label{eq:ode1}
    \sigma_{\rm eff}\ddot{R}=-\frac{2\sigma_{\rm eff}}{R}-\frac{\sigma_{\rm eff}\dot{R}^2}{R}
    +\Delta P-4\eta\frac{\dot{R}}{R}-\dot{\sigma_{\rm eff}}\dot{R}, 
\end{equation}
where $\sigma_{\rm eff}=\kappa\cdot8f_{a}^2m_{a}(t)$ describes the effective domain wall tension  (which does not coincide 
with well known expression $\sigma = 8f_{a}^2m_{a}$ computed in the thin wall approximation), and $\Delta P$ is the  pressure difference  inside and outside the nugget. 
 
 Important element we want to discuss here (in addition to our previous studies)  is related to  the  viscosity term 
$\sim \eta $ which enters eq.   (\ref{eq:ode1}) and effectively describes the friction for the domain wall bubble oscillating 
in   high temperature  plasma. Precisely this term describes  a slow change   of the nugget's size     before the formation is completed and the nugget   assumes its  final form at $T=T_{\rm form}$, see Fig.\ref{phase_diagram}. 
For small  oscillations    the solution of (\ref{eq:ode1}) can be approximated as follows 
\be 
\label{eq:6.R2}
R(t)&=&R_{\rm form}+(R_0-R_{\rm form})e^{-t/\tau}\cos\omega t , \\
 \omega&\sim& {R_{\rm form}^{-1}}, ~~~  \tau\sim\frac{\sigma_{\rm eff}}{2\eta}R_{\rm form}, ~~~ \omega \tau \simeq \frac{\sigma_{\rm eff}}{2\eta}\sim \frac{\Lambda_{\rm QCD}}{m_a} \nonumber 
\ee
which shows the physical meaning of the  frequency $\omega\sim {R_{\rm form}^{-1}}$ and damping time $\tau$. Precisely  parameter $\tau$ describes the time scale when the formation is completed.  By all means it is a highly nontrivial parameter as it represents  a combination of very different scales. Indeed,   the viscosity $\eta$ along any path shown on Fig. \ref{phase_diagram} is always assumes  $\Lambda_{\rm QCD}$ scale (of course it is not known   exactly in different phases);  the axion scale  appears as it enteres through 
$\sigma_{\rm eff}$ and finally, the cosmological scale enters as the formation effectively starts at $T\simeq T_c\simeq 170 $ MeV and must end at $T\simeq T_{\rm form}$ which represents a very long cosmological journey  with typical time scale $t\sim T^{-2} \sim 10^{-4}$ seconds. 

It is a highly nontrivial observation that all these drastically different scales nevertheless lead to a consistent picture.  
Indeed, a typical time for a single oscillation is $\omega^{-1}\sim 10^{-14}s$ for the axion mass $m_a\sim 10^{-4} {\rm eV}$,
while the number of oscillations is very large and of order $\omega\tau\sim 10^{10}$ according to (\ref{eq:6.R2}), see also Appendix ~\ref{appen:evolution}. Therefore, a complete formation of the nugget occurs on a time scale $10^{-4} {\rm s}$ which is precisely the cosmological scale when the temperature drops   to $41~ {\rm MeV}$. This scale is known from completely different arguments related to the estimate of the baryon to photon ratio (\ref{eta}).

 Unfortunately, we could not numerically test this amazing ``conspiracy of scales" in our original   studies  ~\cite{Liang:2016tqc,Ge:2017ttc}. This is because the factor $\omega \tau$ is very large in comparison  with other 
 scales of  the problem.  It is very hard to deal with very  large (or very small)  factors  in numerical computations\footnote{\label{trick}Of course, our case by no means a special in this respect:  it is a common problem in any numerical studies when some parameters assume a parametrical large/small values. It is obviously a case in any numerical studies related to the axion physics because  of a drastic separation of scales,  see e.g.  \cite{Fleury:2015aca,Klaer:2017ond,Gorghetto:2018myk}.}.
 This is precisely the reason why in numerical analysis in ~\cite{Liang:2016tqc,Ge:2017ttc}
 the viscosity term 
$\sim\eta$ was  artificially enlarged $\sim10^8$ times to make eq.~(\ref{eq:ode1}) numerically solvable, which is a conventional technical trick, see footnote \ref{trick}.
%This is because the viscosity term $\eta\sim 1$ (in the QCD unit) is very small compared with other 
%terms in eq.~(\ref{eq:ode1}) with $\eta/\sigma\sim10^{-10}$. As a consequence, a compromise approach 
%is taken in Refs.~\cite{Liang:2016tqc,Ge:2017ttc} by enlarging the term $\eta$ to an unphysical value 
%to overcome this multi-scale problem. 

\begin{figure}
    \centering
    \includegraphics[width=\linewidth]{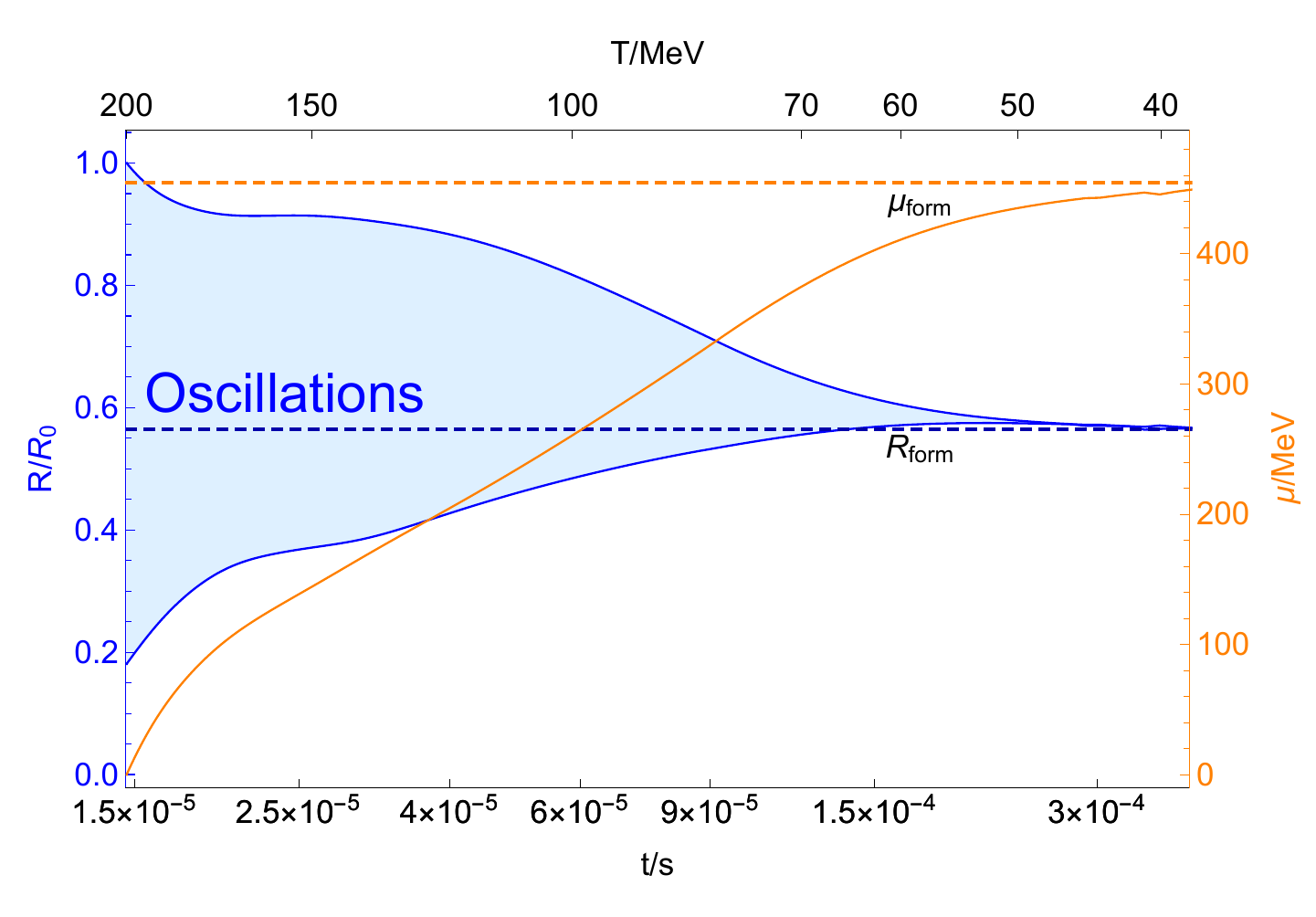}
    \caption{Numerical result of the nugget evolution. The two solid blue lines represents respectively 
    the upper envelope and the lower envelope of $R$ oscillations. The shaded light blue region 
    represents the numerous oscillations. The solid orange line represents the lower envelope of $\mu$ 
    oscillations (we did not show the upper envelope and shaded region for $\mu$ oscillations to make 
    the picture more clear). The dashed blue line and dashed orange line represents respectively 
    $R_{\rm form}$ and $\mu_{\rm form}$ using simple analytical arguments as expressed   by eqs.~(\ref{eq:R_1}) and (\ref{eq:R_2}). We see 
    that they match the numerical result of nugget evolution pretty well.}
    \label{fig:RTE}
\end{figure}

It was one of the goals of the present work to overcome this technical difficulty 
 by adopting  a new numerical method coined as 
\textit{envelope-following} method which can solve our system successfully with the viscosity term 
keeping its real physical magnitude $\eta\sim \Lambda_{\rm QCD}^3$ when parameter $\omega\tau\sim 10^{10}$ assumes its very large physical values.  We describe the    method and present  the numerical analysis   in 
Appendix~\ref{appen:evolution}.  Here we only summarize the basic results of these  studies which  
   confirm the  main  features of the AQN model,  
see Fig.\ref{fig:RTE}: \\1. the nugget completes its evolution by oscillating numerous number of times $\omega\tau\sim 10^{10}$ before it assumes its final configuration with size $R_{\rm form}$ at $T_{\rm form}\approx 40 ~{\rm MeV}$.
Therefore, the ``conspiracy of scales" phenomenon mentioned above has been explicitly tested; \\
2.  the chemical potential inside the nugget indeed assumes a sufficiently large value  $\mu_{\rm form}\gtrsim 450$~MeV during this long evolution. This magnitude  is consistent with  formation of a 
  CS phase. Therefore, the    original assumption on CS phase which was used in construction of the nugget is justified a posteriori.  

\section{Baryon charge distribution}\label{sec:distribution}
The main  goal of this section is to  calculate the baryon charge distribution of nuggets in the AQN scenario  and 
compare it to the observational constraints listed in section \ref{sec:Constraints}. We start in subsection \ref{basics}
with  formulating of the basic idea of the computations. In two  subsections \ref{subsec:R0} and \ref{subsec:T0} which follow we study initial size and temperature distributions correspondingly.  Finally, in subsection \ref{subsec:plotsNB} we formulate our  main results on the nugget's distribution $dN/dB$.

\subsection{The basic idea of computations}\label{basics}
In the present section we need a relation between initial size of the nugget $R_0$ formed  at the temperature $T_0$ and its total baryon charge 
when the stage of formation is completed. 
The desired relation reads,  
\begin{equation}\label{eq:Brelation}
    B\simeq  K\cdot R_{0}^3 T_{0}^3, ~~~~ K\equiv \frac{\pi^2}{27\sqrt{6}}g^{\rm in}
\end{equation}
  see Appendix \ref{appen:evolution} with  the detail analysis regarding   relation (\ref{eq:Brelation}). 
This  relation implies  that the total baryon charge $B$ of a stable nugget is completely determined by the 
initial size $R_{0}$ and the initial temperature $T_{0}$ of the closed domain wall such that 
  $B\propto(R_{0}T_{0})^3$.

Eq.~(\ref{eq:Brelation}) tells us that closed domain walls with different initial radii and temperatures 
will eventually carry different baryonic charges $B$. Since the closed domain walls can form with 
different initial radii and at different temperatures ($T_{c}\lesssim T_{0}\lesssim T_{\rm osc}$), we may map these initial conditions onto a baryon charge distribution of the nuggets $dN/dB$. 

\exclude{
In this context it 
is largely the distribution of those AQN composed of antiquarks that is important as the possibility 
of their annihilation with visible matter means they dominate the energy budget of any observable
quantities. However, both nuggets and anti-nuggets form from the initial closed axion domain walls 
so they should have approximately the same distribution on the baryon charge.
The relation~(\ref{eq:Brelation}) 
is the first step for us to calculate the distribution, for it points out that the baryon charge carried by 
a single nugget is determined from its initial conditions. With this knowledge, we are going to discuss 
the distributions of $R_{0}$ and $T_{0}$ in the next section, from which we can get the baryon 
charge distribution.
}

According to eq.~(\ref{eq:Brelation}), the baryon charge distribution ($dN/dB$) of stable nuggets 
can be obtained from the initial size ($R_{0}$) and initial temperature ($T_{0}$) distributions of 
the closed axion domain walls which form between $T_{\rm osc}$ and $T_{c}$ as the initial stage of the 
nuggets. We start with the following equation
\begin{equation}\label{eq:dN}
    dN=N_{0}\cdot P\cdot f(R_{0},T_{0})\cdot dR_{0}dT_{0},
\end{equation}
where $dN$ is the number of closed domain walls with the initial radius in the range 
$(R_{0},R_{0}+dR_{0})$ and the initial temperature in the range $(T_{0},T_{0}+dT_{0})$; $f(R_{0},T_{0})$ 
is  two parametrical distribution function which represents the probability density of a closed domain wall with 
$R_{0}$ and $T_{0}$ in the above ranges. The factor $N_{0}$ is the total number of closed bubbles that form in the early Universe when 
$T_{a}\lesssim T_{0}\lesssim T_{c}$, while   $P$ is a normalization factor to make the probability density 
$f(R_{0},T_{0})$ normalized to one, i.e. 
\begin{equation}\label{eq:normalization}
    \iint P\cdot f(R_{0},T_{0})\cdot dR_{0}dT_{0}=1.
\end{equation}
%which means $P$ is calculable in principle once $f(R_{0},T_{0})$ is known.  
The main goal of this section is to develop a technique which allows to compute  
$f(R_{0},T_{0})$. 

To simplify our analysis we assume that all initial closed domain walls will eventually  become the stable nuggets. 
We clarify this assumption later in the text when we compare the prediction of our construction with 
observational constraints. As the next step we use  
the relations (\ref{eq:Brelation}) and (\ref{eq:dN}), to represent the  number of stable nuggets with the baryon 
charge \textit{less} than $B$ as follows
\begin{equation}\label{eq:NB}
N(B)=\iint_{K\cdot R_{0}^3 T_{0}^3\leq B}N_{0}\cdot P\cdot f(R_{0},T_{0})\cdot dR_{0}dT_{0},
\end{equation}
where $K\cdot R_{0}^3 T_{0}^3\leq B$ constraints the parametrical space of  the integration.

From eq.~(\ref{eq:NB}), we can further calculate the baryon charge distribution $dN(B)/dB$ which is the main topic of this section. Obviously, 
the  distribution $f(R_{0},T_{0})$ which   depends on $T_0$ and $R_0$ in a very nontrivial way plays a crucial rule in our calculations of the $dN(B)/dB$ distribution. 
  The study of the function $f(R_{0},T_{0})$ can be approximately separated into two distinct pieces: one part 
describes  the $R_{0}$- dependence,  while  the   $T_{0}$ distribution can be incorporated separately. Next two subsections  are devoted to analysis  of    these two different elements of the main problem.  

\subsection{Initial size distribution}\label{subsec:R0}
As discussed in section~\ref{sec:RTE}, in the AQN model the $N_{\rm DW}=1$ domain walls 
are the topological defects with the axion field $\theta$ interpolating between $k=0$ ($\theta=0$) and 
$k=1$ ($\theta=2\pi$) branches. Although $k=0$ and $k=1$ branches correspond to the same 
unique physical vacuum,  they effectively act as two different vacua with the same energy.
 The domain walls can  interpolate between these (physically identical but topologically distinct) vacua, similar to a model with $V(\theta)\sim \cos\theta$ potential, when $\theta=0$ and $\theta=2\pi$ correspond to one and the same   physical vacuum.  Therefore, the $N_{\rm DW}=1$ axion domain walls in  this scenario can be treated as $Z_{2}$ domain walls  which greatly  simplifies  the computations. 

The closed $Z_{2}$ domain walls 
have been observed in the simulations of $Z_{2}$-wall system~\cite{vachaspati1984formation}. In 
our case, it means that closed $N_{\rm DW}=1$ axion domain walls can form, which are the 
sources of stable nuggets as we discussed in section~\ref{sec:RTE}.
Furthermore, this analogy will provide us with more useful information about the initial size distribution 
of these closed bubbles. The Ref.~\cite{vachaspati1984formation} points out that the probability of 
forming a closed $Z_{2}$ domain wall with the initial radius $R_{0}\gg\xi$ (where $\xi$ is the correlation 
length of the topological defects) exponentially suppressed,  $\sim\exp{(-R_{0}^2/\xi^2)}$. The procedure 
in Ref.~\cite{vachaspati1984formation} to derive this relation is briefly reiterated as follows. 

To simulate the $Z_{2}$ system in three dimensions, we first divide a big cubic volume into many 
small cubic cells, each of which has the length $\xi$. Then to each cell a number a number $+1$ or 
$-1$ is assigned at random with equal probability $p=0.5$. This is the simulation of the phenomenon 
that different patches (with volume $\sim \xi^3$) of the space during the phase transition will settle 
randomly with equal probability in one of the two vacua ($\theta=0$ and $\theta=2\pi$ in the case of 
$N_{\rm DW}=1$ axion domain walls). The domain walls lie on the boundaries between cells of 
opposite sign. Two neighbouring cells are connected if they have the same sign. Many connected cells 
can form a cluster with the same sign. The size $s$ of a cluster is defined as the number of cells in 
the cluster. We then can look for the size distribution of $+1$-clusters (Of course, the size of 
$-1$-clusters will follow the same distribution). It turns out that this is a typical problem of the 
percolation theory, which deals with the statistics of the clusters at different values of $p$. See 
Refs.~\cite{STAUFFER19791,stauffer2014introduction} for a review of the percolation theory\footnote{In 
percolation theory, there is a percolation threshold $p_{c}$, at which an infinite cluster first appears 
in an infinite lattice. $p_{c}=0.31$ in three dimensions for a cubic lattice. In our case where the probability 
of a cell picking $+1$ is $p=0.5$, we have one infinite $+1$-cluster ($p>p_{c}$) and one infinite 
$-1$-cluster ($1-p>p_{c}$). In the language of domain walls, it can be interpreted as the system being 
dominated by one infinite wall of very complicated topology~\cite{vachaspati1984formation}. In addition 
to this infinite domain wall, there are some closed domain walls (finite clusters) and they satisfy the 
size distribution~(\ref{eq:ns}). The structure and the dynamics of the infinite domain wall is less important 
for our present work which is focussed on the closed domain walls.}. 

In our case where $p=0.5$ in three dimensions, the size distribution of the finite clusters is known from the percolation 
theory~\cite{STAUFFER19791}, which has the following expression
\begin{equation}\label{eq:ns}
    n_{s}\propto s^{-\tau}\exp{(-\lambda s^{2/3})},
\end{equation}
where $n_{s}$ is the number density of finite clusters as a function of the cluster size $s$ (the number 
of the cells inside a cluster). Although the distribution~(\ref{eq:ns}) is derived for large clusters 
$s\gg1$~\cite{STAUFFER19791}, it turns out that this relation can be extrapolated down to $s=1$ as 
a very good approximation~\cite{BAUCHSPIESS1978567}. As a consequence, we adopt 
eq.~(\ref{eq:ns}) for the whole spectrum $s\geq1$ for further calculations. The two coefficients $\tau$ 
and $\lambda$ are $p$-dependent. According to the Ref.~\cite{BAUCHSPIESS1978567}, $\lambda$ 
has a typical value $\sim 10$ and $\tau$ ranges from $1.5$ to $2.2$ based on the three dimensional 
lattice simulations. Discussing the exact values of $\tau$ and $\lambda$ at $p=0.5$ is beyond the 
scope of this work. Instead, we simply adopt $\lambda=10$ and $\tau=2$ for further calculations\footnote{$\lambda$ can also be calculated using the relation $\lambda^{-1}\simeq \left|p-p_{c}\right|^{-1/\sigma_{p}}$ where $\lambda^{-1}$ is the crossover size (see e.g. Refs.~\cite{STAUFFER19791,isichenko1992percolation,grinchuk2002large}).  This relation is valid for $\left|p-p_{c}\right|\ll 1$. The parameter $\sigma_{p}=0.45$ in 3D~\cite{stauffer2014introduction}. We then get $\lambda\approx0.025$ for $\left|p-p_{c}\right|\ll 1$ satisfied. In addition, $\tau=-1/9$ for $p>p_{c}$ is obtained in a field theoretical formulation of percolation problem~\cite{lubensky1981cluster}. However, the exact values of $\lambda$ and $\tau$ are not important for us, since they do not affect the slope of the distribution $dN(B)/dB$ as we will see in section~\ref{subsec:plotsNB}.}. 
However, as we will see, the shape of the baryon charge distribution $dN(B)/dB$ of nuggets is not sensitive to 
the precise numerical values of $\tau$ and $\lambda$. 

The result~(\ref{eq:ns}) can be translated into the language of domain walls straightforwardly: 
The probability of forming a closed bubble with radius $R_{0}$ decreases exponentially when 
$R_{0}$ increases, which can be formally expressed as 
\begin{equation}\label{eq:R0}
    \frac{dN}{dR_{0}}\propto \xi^{-1}\left(\frac{R_{0}}{\xi}\right)^{2-3\tau}\cdot
    \exp{\left[-\lambda\left(\frac{R_{0}}{\xi}\right)^{2}\right]}.
\end{equation}
To derive this distribution as a function of $R_{0}$ from eq.~(\ref{eq:ns}), we used the relations 
$s\simeq R_{0}^3/\xi^3$ and $n_{s}=\frac{1}{V}\frac{dN}{ds}$ where we get rid of the simulation 
volume $V$ (a constant) in eq.~(\ref{eq:R0}). The parameter $\xi$ is the correlation length of topological 
defects as mentioned above, which is also set as the length of a single cell. The smallest cluster is a 
cell ($s\geq 1$) implying that the lowest bound of the radius of closed bubbles is $R_{0}\gtrsim\xi$. 
Since the relation~(\ref{eq:ns}) is applicable for all finite clusters $s\geq1$ as mentioned above, we 
adopt eq.~(\ref{eq:R0}) as the size distribution of all closed bubbles $R_{0}\gtrsim\xi$.

It is very instructive to consider an oversimplified case where there is no initial temperature distribution. It can be 
realized  if all the closed bubbles form at the same moment (at the same temperature). In this case 
the  distribution $f(R_{0}, T_{0})$ does not depend on $T_0$ and, according to 
eq.~(\ref{eq:R0}),  can be written as 
$f(R_{0})=dN/dR_{0}\propto \xi^{-1}(R_{0}/\xi)^{2-3\tau}\exp{[-\lambda(R_{0}/\xi)^2]}$. Using eq.~(\ref{eq:Brelation}) this $R_0$ dependence can be translated into $ \frac{dN}{dB}$ distribution:  
\be
\label{eq:simplified_model}
    \frac{dN}{dB}&=&\frac{dN}{dR_{0}}\frac{dR_{0}}{dB}\\
    &\propto& \frac{1}{B_{\rm min}}\left(\frac{B}{B_{\rm min}}\right)^{-\tau}
    \exp{\left[-\lambda\left(\frac{B}{B_{\rm min}}\right)^{\frac{2}{3}}\right]}, \nonumber
\ee
where $B_{\rm min}\equiv K\cdot \xi^3 T_{0}^3$. In this oversimplified model where there is 
no $T_{0}$ distribution, we find $dN/dB$ is  greatly 
suppressed by the exponential factor $\sim\exp{[-\lambda(B/B_{\rm min})^{2/3}]}$. 
This essentially would imply  that the distribution is strongly peaked at $B\approx B_{\rm min}$, while larger bubbles are strongly suppressed. 

As we discuss in next subsection the $T_0$-dependence drastically and qualitatively  changes this simplified picture.
The key element is that  the closed bubbles initially 
form at different temperatures between $T_{\rm osc}$ and $T_{c}$ as discussed above.
The correlation length $\xi\sim m_a^{-1}$ which is inversely proportional to the axion mass $m_a$
  drastically changes during this evolution because of the dramatic changes of the axion mass  in this interval. 

 These profound  changes   completely modify  the basic features of  the  distribution function $f(R_{0},T_{0})$, which is the subject of  the following subsection. As we shall  see below, the  baryon charge distribution    satisfies 
a power-law   $dN/dB\propto B^{-\alpha}$ when $T_0$ dependence is properly incorporated, rather than follows the exponential behaviour (\ref{eq:simplified_model}). This   power law is consistent with  parametrization (\ref{distribution}) which has been postulated to fit the observations. Furthermore, power-law   $dN/dB\propto B^{-\alpha}$ behaviour as we discuss below  is not very sensitive to  the parameters  of coefficients $\tau$ and $\lambda$, and therefore, represents a very robust consequence of the framework. 

\subsection{Initial temperature distribution and the correlation length $\xi(T)$}\label{subsec:T0}
As we discussed in section~\ref{sec:RTE}, the closed axion domain walls could form anywhere between 
$T_{\rm osc}$ and $T_{\rm c}$, see Fig. \ref{phase_diagram} to view the phase diagram corresponding this evolution. It is hard to calculate the exact $T_{0}$ distribution. It is known, though, that normally the temperature dependence enters implicitly through the correlation length $\xi(T)$ which is highly sensitive to the temperature.

 To account for the corresponding modifications 
we adopt a conventional assumption that the correlation length is a few times  the domain wall width 
$\xi(T)\sim m_{a}^{-1}(T)$. The axion mass is known to be a temperature-dependent function before 
it reaches its asymptotic value near $T_{c}$  because it  is  proportional to the 
topological susceptibility. At sufficiently high temperature $T\gg T_c$ one can use the instanton liquid model 
\cite{Wantz:2009mi,Wantz:2009it} to estimate the power law $m_{a}(T)\propto T^{-\beta}$. When the temperature 
is close to $T\simeq T_c$ one should use the lattice results  to account for a  proper temperature scaling of the axion mass. 

The recent lattice QCD result shows\footnote{The 
Ref.~\cite{Borsanyi:2016ksw} does not show the value of $\beta$ explicitly, but provides the related 
data in the Supplement Information. We get $\beta=3.925$ by fitting the data provided. The lattice results are, in fact,  consistent  with   analytical models~\cite{Wantz:2009mi,Wantz:2009it},  see also the Appendix~\ref{appen:evolution} for additional details.} that 
$m_{a}(T)\propto T^{-\beta}$ with $\beta=3.925$ just above  $T_{c}$~\cite{Borsanyi:2016ksw}. We then can approximate  the correlation length  in the entire interval as
\begin{equation}\label{eq:xiT}
    \xi(T_{0})=\xi_{\rm min}\cdot\left(\frac{T_{0}}{T_{c}}\right)^{\beta},~~~~~T_{c}\lesssim T_{0}\lesssim T_{\rm osc}
\end{equation}
where $\xi_{\rm min}\equiv\xi(T_{0}=T_{c})$ is the minimal correlation length. The same $\xi_{\rm min}$ also serves as 
the minimal radius that closed bubbles could have because $R\gtrsim\xi$. 

In what follows we also assume   the   following   
simple model to account for the temperature variation of the $ {dN}/{dT_{0}}$ distribution\footnote{One subtlety is that the effect of the expansion of the Universe 
between $T_{\rm osc}$ and $T_{c}$ is also included in the model (\ref{eq:T0distribution}), since $N$ is defined 
as the number of closed domain walls rather than the number density.},
\begin{equation}\label{eq:T0distribution}
    \frac{dN}{dT_{0}}\propto \frac{1}{T_{c}}\left[\frac{\xi(T_{0})}{\xi(T_{c})}\right]^{\delta}\propto
    \frac{1}{T_{c}}\left[\frac{T_{0}}{T_{c}}\right]^{\beta\delta},
    \end{equation}
where $\delta$ is a free parameter adjustable to shape different $T_{0}$ distributions. This 
parameterization has the advantage of producing a simple final expression for the baryon number 
distribution while still capturing the essentials of the temperature dependance. The constant 
$1/T_{c}$ has no special physical meaning but is introduced to balance the units of the right-hand side 
and the left-hand side of the relation. Perhaps the simplest case is $\delta=0$, in which case $T_{0}$ 
is a uniformly distributed, i.e. the probability of forming nuggets is uniform between $T_{\rm osc}$ and $T_{c}$.
One should emphasize that $\delta=0$ case is still not reduced to  the oversimplified example mentioned at the end of the previous subsection.  This is because the temperature dependence  explicitly enters through (\ref{eq:T0distribution}), but it  also enters implicitly through the temperature dependence of the correlation length $\xi(T)$ in formula (\ref{eq:R0}). 

For positive $\delta>0$, the nuggets tend to  form close to the point $T_{\rm osc}$,  while for negative  $\delta<0$, nuggets tend to form   when the tilt becomes much more pronounced  close to 
the QCD transition temperature $T_{c}$.  Sufficiently large numerical value of $|\delta|> 1$ with any sign
corresponds to a very sharp, almost explosive for $|\delta|\gg 1$,   increase of the probability for the axion bubble formation at $T\simeq T_{\rm osc}$ or at  $T \simeq T_c$ depending on sign of $\delta$. At the same time  $|\delta|\sim  0$
corresponds to a very smooth behaviour in the entire temperature interval  (\ref{eq:T0distribution}).
 We, of course, do not know any 
properties of  the distribution (\ref{eq:T0distribution}) in strongly coupled QCD when $\theta\neq0$. Therefore, we 
proceed with our computations with arbitrary $\delta$ and make comments on the obtained properties  of the  baryon distribution  $dN/dB$ as a function of unknown parameter $\delta$ in next subsection \ref{subsec:plotsNB}.

Combining the $T_{0}$ distribution (\ref{eq:T0distribution}) with the $R_{0}$ distribution (\ref{eq:R0}), 
\exclude{
one arrives to the two parametric  distribution $f(R_{0},T_{0})$ which represents  the product of these two distributions
\be
\label{eq:f0}
    f(R_{0},T_{0})&=&\frac{1}{\xi T_c}\left(\frac{T_{0}}{T_{c}}\right)^{\beta\delta}\cdot \left(\frac{R_{0}}{\xi}\right)^{2-3\tau}
   \cdot \exp{\left[-\lambda\left(\frac{R_{0}}{\xi}\right)^{2}\right]}, \nonumber  \\
    &&~T_{c}\lesssim T_{0}\lesssim T_{\rm osc},~~~~~R_{0}\gtrsim\xi.
\ee
}
and substituting eq.~(\ref{eq:xiT}) 
into eq.~(\ref{eq:R0}), we  arrive to the following two-parametric distribution function,
\be
\label{eq:f}
    f(R_{0},T_{0})&=&\frac{1}{\xi_{\rm min}T_{c}}\cdot \left(\frac{T_{0}}{T_{c}}\right)^{3\beta(\tau-1)+\beta\delta}
    \cdot\left(\frac{R_{0}}{\xi_{\rm min}}\right)^{2-3\tau} \nonumber \\
    &\times&\exp{\left[-\lambda\left(\frac{R_{0}}{\xi_{\rm min}}\right)^{2}
    \left(\frac{T_{c}}{T_{0}}\right)^{2\beta}\right]},\\
    &&T_{c}\lesssim T_{0}\lesssim T_{\rm osc},~~~~R_{0}\gtrsim\xi(T_{0}). \nonumber
\ee
Notice that here we use ``$=$'' rather than ``$\propto$''. This is because we have an extra factor $P$ 
in eq.~(\ref{eq:dN}) which serves as the normalization factor, and the constant multipliers in 
$f(R_{0},T_{0})$ can be collected and included into $P$.

With this expression for $ f(R_{0},T_{0})$ and   basic  eq.~(\ref{eq:NB}) we can now  proceed 
with calculation of  the  baryon charge  
distribution $dN/dB$. The corresponding  results will be discussed in the next subsection. 

\subsection{The  $dN/dB$ distribution. Results. }\label{subsec:plotsNB}
Substituting eq.~(\ref{eq:f}) into eq.~(\ref{eq:NB}), one can explicitly compute the  function $N(B)$ and   the 
distribution $dN/dB$.   In what follows it is convenient to introduce the following dimensionless 
variables: the baryon charge $b=B/B_{\rm min}$ of the nugget measured from  its minimum value  $B_{\rm min}=K\xi_{\rm min}^3T_{c}^3$;  the relative size $r=R_{0}/\xi_{\rm min}$ of the nugget  measured from its minimum size 
$\xi_{\rm min}$; the relative temperature $u=T_{0}/T_{c}$ during formation evaluation in units of  $T_c$.

 In terms of these dimensionless variables the desired  distribution $ {dN}/{dB}$ can be represented as follows 
 \be
 \label{eq:f1A_copy}
 &&   \frac{dN}{dB}=\frac{N_{0}P}{3B_{\rm min}}\cdot \left(\frac{1}{b}\right)^{\tau}  \\ 
 &\times&\int_{1}^{b^{\frac{1}{3(\beta+1)}} }du~
    \left[u^{3(\beta+1)(\tau-1)+\beta\delta} {\rm e}^{-\lambda b^{ {2}/{3}} u^{-2(\beta+1)}} \right],\nonumber
\ee
 see Appendix~\ref{appen:NB} with all technical  details. 
 
 One can easily estimate the integral (\ref{eq:f1A_copy}) by observing that it is saturated for very large $b\gg1$ by
 $u_{\rm sat}$   of order   
 \be
 \label{u}
 u_{\rm sat}\sim \left[\lambda b^{2/3}\right]^{\frac{1}{2(\beta+1)}}\sim b^{\frac{1}{3(\beta+1)}} , ~~~~~ b\gg 1
 \ee
 when the exponential factor in  (\ref{eq:f1A_copy}) assumes a value of  order  one.  Substituting the expression back to eq. (\ref{eq:f1A_copy}) 
 one arrives  to the following   asymptotical behaviour for the distribution  
 \begin{equation}
\label{eq:slope_prop}
    \frac{dN}{dB}\propto B^{-\alpha},~~B\gg B_{\rm min},
\end{equation}
where the final  result  is  expressed  in  terms of the physical baryon charge  $B$ rather than in terms of the dimensionless parameter $b$.
Parameter $\alpha$ here is defined precisely in the same way as it is defined in the observational fitting formula (\ref{distribution}). 

The exponent 
$\alpha $ entering (\ref{eq:slope_prop}) 
can be approximated in the  limit $B\gg B_{\rm min}$  as follows
\begin{equation}
\label{eq:alpha_nugget}
    \alpha \approx 1-\frac{\beta\delta+1}{3(\beta+1)} \sim 1-\frac{\delta}{3}, 
\end{equation}
where in the last step we ignored the factors of order one in comparison with known (and very large) value of $\beta\simeq 4$ to simplify qualitative discussions below. The approximate analytical formula   (\ref{eq:slope_prop}) at very large $B\gg B_{\rm min}$ is in perfect agreement with numerical analysis presented in  Appendix~\ref{appen:NB}. 

The behaviour (\ref{eq:slope_prop}) is amazingly simple and profoundly important result. Indeed, it shows that 
the exponential suppression is replaced by the algebraic decay (\ref{eq:slope_prop}) which is consistent with  observational fitting formula 
(\ref{distribution}). The ``technical" explanation for  this to happen is that the integral (\ref{eq:f1A_copy}) is saturated by $u_{\rm sat}$ 
 when the exponential factor in  (\ref{eq:f1A_copy}) assumes a value of  order  one. In terms of the physical parameters it  is related to the fact  
that exponential suppression (\ref{eq:f}) due to the large size $R_0$ is effectively removed  by a strong temperature dependence with very large beta function $\beta$. Integration  over entire  temperature interval   eventually leads to the algebraic decay (\ref{eq:slope_prop}). 

Another important property of the expression (\ref{eq:slope_prop})    is that the final result for the slope (\ref{eq:alpha_nugget})
is not very sensitive to the parameters $\lambda$ and $\tau$. The total normalization factor of course is very sensitive to these parameters as discussed in  Appendix~\ref{appen:NB}. It is also not very sensitive to the well known parameter $\beta\approx 4$ as long as it is relatively large.  The slope $\alpha $ is mostly determined by 
$\delta$ which may have any sign and effectively describes the temperature interval where the bubbles are produced with the highest efficiency.  
 The fitting models (\ref{distribution}) based on observations which were discussed in Section \ref{AQN} can be reproduced with  negative $\delta<0$. The negative sign of $\delta$   as we previously mentioned  corresponds to the 
 preference of the bubble formation close to $T_c$ where the axion potential tilt becomes much more pronounced. Furthermore, a model with $\alpha \simeq  2$ corresponds to $\delta\simeq -3$ (strongly peaked at $T\simeq T_c$), while another model with $\alpha \simeq 1.2$ corresponds to a more smooth distribution of  ${dN}/{dB}$ over the entire temperature interval with $\delta\simeq -1$ corresponding to mild preference of the bubble formation at $T\simeq T_c$.

The last comment we want to make  is about the largest possible size of nuggets. According to percolation theory, there is 
no upper limit on the size of finite clusters (closed domain walls). However, the shape of large clusters 
may not be perfectly spherical (in 3D) while our computations are based on assumption of exact 
 spherical symmetry    of the formed bubbles. 
Furthermore,   the radius for non-symmetric bubbles  is defined in  average sense for large closed clusters, see e.g.~
\cite{stauffer2014introduction} for more details. The deviation from the ideal spherical shape makes 
the large collapsing closed domain walls to  fragment with high probability  into smaller pieces and thus could 
significantly suppress the possibility of forming large nuggets\footnote{The ref.~\cite{vachaspati2017lunar} 
presents a similar argument when the author discusses the possibility of domain wall membranes 
(e.g. closed domain walls) collapsing into black holes.}. The detailed calculations of the suppression 
effect from the irregular shape for large clusters is hard to carry out and also well beyond the scope of 
the present work. However, we may introduce a cutoff $B_{\rm cut}$ to roughly account for this 
extra suppression. Above $B_{\rm cut}$, no nuggets can form from the collapse of closed axion 
domain walls. This parameter turns out to be useful when we later calculate the total number of 
nuggets.

We conclude  this section  with the following remark. The main result of  our analysis is  expressed as  
 (\ref{eq:slope_prop}) with the slope (\ref{eq:alpha_nugget}). This formula  represents the baryon charge distribution 
 immediately after the formation period is complete when  the baryon to photon ratio $\eta$ assumes its present value (\ref{eta}). This ``primordial" distribution of the nuggets is the subject of a long evolution in hot plasma
 which may modify the properties of $dN/dB$.  This problem of ``survival" of the primordial nuggets  is the subject of the next section.  

\section{Survival of the Primordial Distribution}\label{sec:survival}
   
After the AQN have formed at $T\approx 40 $ MeV the process of ``charge separation"   
is essentially complete and the 
plasma surrounding the nuggets contains exclusively protons, neutrons, electrons and positrons. A nugget 
composed of matter will gradually collect electrons into its electrosphere as the plasma cools 
but apart from this will essentially remain in its initial form. The surface layer of electrons contribute 
negligibly to the total mass so that the distribution of nugget masses remains essentially identical 
to the primordial distribution discussed above. However, the AQN composed of antimatter, which are 
present in larger numbers, will be subject to a much more complicated evolution. The details 
of this process will be laid out below, but we first give a general overview of the evolution of the 
antimatter AQN mass distribution.

Initially the plasma surrounding the AQNs is dominated by electrons and positrons which are roughly 
as abundant as the photons, i.e.  $n_e\simeq n_{e^+}\simeq n_{\gamma}\sim T^3$. During this phase the electrosphere captures positrons into those states 
for which the binding energy is above the plasma temperature and expands similar to the case of 
a matter nugget. However, once the temperature drops below the electron mass  $T\leq m_e$ the electrosphere 
can no longer capture free positrons at a rate sufficient to compensate for annihilations. 
Below 
this temperature the electrosphere will begin to capture free protons which, if they stay bound to the 
nugget for a sufficient period of time, will eventually annihilate with the central quark matter. 

The process of capturing protons become much more pronounced after the temperature drops to $T\approx 20$ keV
when the dominant portion of the positrons in the plasma get annihilated, while the number density of electrons and protons become equal, 
i.e. $n_e\approx n_p\sim \eta T^3$. However, even at this temperature, as we discuss below,  only 
a very tiny  portion of the AQN's baryon charge 
will be annihilated, such that mass distribution still remains essentially unaffected by the 
unfriendly environment in form of the hot plasma.

Finally, 
after recombination at $T\leq 1$ eV the surrounding matter is largely neutral and at much lower densities. During 
this time matter (primarily in the form of neutral hydrogen) will continue to collide with the antimatter 
AQN with some probability of annihilation but at a relatively low rate. The rare events of annihilation during the present time 
lead to a number of observable effects as reviewed in Section \ref{sec:Constraints}.
 
At each phase of evolution the scattering rate of baryons on the nugget (and thus the probability of 
an annihilation) scales with the cross-section of the nugget. This is  at least approximately true 
even in the case where long range electrical effects must be considered as the nuggets' electrical 
charge is itself a surface effect. As such any change in the mass distribution should be expected 
to show a $\Delta M/M \sim \sigma/M \sim B^{-1/3}$ behaviour. 

The following sections will trace the evolution of the mass distribution from formation to the 
present day. Specifically, in next section  \ref{ssec:PreBBN} we study the evolution of the nuggets 
in very hot plasma 
before BBN epoch.    In section   \ref{low-T} we analyze the AQN evolution before the recombination, while in section \ref{recombination} we the evolution of the nuggets after the recombination including the period of the galaxy formation.
Finally, in section \ref{present}  
   we study the evolution of the AQNs at  present day Universe.  
 We will demonstrate that, for a range of physically interesting parameters, the initial 
population of AQN will survive until the present day as a population consistent with all 
observational constraints and with the parameter space allowed for the axion mass and the AQN's baryon charge $B$ as discussed in Section \ref{sec:Constraints}.

\section{Pre-BBN Evolution}\label{ssec:PreBBN}
The AQN complete formation and settle into a stable colour superconducting phase at a 
temperature of approximately $T_{\rm form} \approx 40$ MeV, see Fig.\ref{phase_diagram}. Once this transition is complete the 
AQN will cease accreting mass and annihilation with the free baryons in the plasma will become 
the dominant process\footnote{The plasma already possesses the required baryon asymmetry at this 
time so only the antimatter AQN will be subject to annihilation while the AQN made of matter experience 
only elastic scattering.}. However, annihilation between an energetic free baryon and the quark content 
of the AQN is highly nontrivial process as we discuss below.  

\exclude{For annihilation to be successful the incident baryon must 
have sufficient time for adjustment its quark's   wave function  with the complex non-baryonic di-quark state of the 
colour superconductor. This will result in a substantial suppression of the rate at which colliding 
baryonic matter is able to annihilate as we discuss below. 
}

We start with the estimates of the collision rate in pre-BBN epoch. 
The corresponding  rate between an AQN and the baryons of the surrounding plasma is,
\begin{equation}
\label{eq:hot_rate}
\Gamma_{\rm col} = 4\pi R^2 n_B v_B = 4\pi R^2 \frac{2\zeta (3)}{\pi^2} 
\eta \left( \frac{T}{\hbar c}\right)^3  \sqrt{\frac{2T}{m_p}} c
\end{equation}
%where the baryon to photon ratio $\eta$ is slightly larger than it's present value as the electrons 
%and positrons are effectively massless at this time. 
where the baryon number density in plasma $n_B$ can be approximated  as $n_B\sim \eta T^3$. 
The total number of collisions during this 
time period is saturated by the highest temperature $T_{\rm form}\simeq 40~{\rm MeV} $  and can be estimated as follows 
\begin{equation}
\label{eq:hot_col}
 \begin{aligned}
N_{\rm col} = \int dt ~\Gamma = \int_0^{T_{\rm form}} dT~ \frac{dt}{dT}~ \Gamma_{\rm col} \\
\approx 3\times 10^{25} \left(\frac{T_{\rm form}}{40~{\rm{MeV}}}\right)^{1.5} 
\left(\frac{R}{10^{-5}{\rm{cm}}}\right)^2 
\end{aligned}
\end{equation} 
where we have used the relation $t\sim T^{-2}$ to change to a temperature integration. 

While the number of collisions (\ref{eq:hot_col}) is comparable with the total baryon charge $B$ of a nugget,
the probability of annihilation is quiet small. Instead, 
the most likely interaction of any incident matter with 
the nugget is total reflection due to a number of reasons: the sharp boundary between hadronic  and  CS phases such that  only a very small fraction $\kappa (T)\ll 1$ of collisions represented by (\ref{eq:hot_col})
will result in an annihilation.  We refer to Appendix \ref{kappa} for order of magnitude estimates supporting the main claim 
that $\kappa (T)\ll 1$. A more precise vale is not essential for our arguments which follow. 

So long as the electrons and positrons remain relativistic (and thus 
present in numbers comparable to the photons) all long range interactions are effectively screened 
and the cross section appearing in expression (\ref{eq:hot_rate}) is purely the physical size of the 
AQN. As such the estimate (\ref{eq:hot_col}) holds  until much lower temperatures when 
the positrons have fully annihilated (which approximately occurs at $T\approx 20$ keV) and longer range interactions become possible. 
The estimate (\ref{eq:hot_col}) implies that the number of annihilation events does not modify 
the primordial spectrum of the AQNs discussed in Section  \ref{subsec:plotsNB} because the relative number of annihilation events is very small, i.e. $(\kappa N_{\rm col})/B\sim \kappa\ll 1$.

While the baryon charge annihilation events are strongly suppressed by the factor $\kappa\ll 1$ the $e^+e^-$ annihilation events   involving  particles from AQN's electrosphere are much more numerous and unsuppressed. One may therefore wonder if the energy injection by these annihilation events may impact the conventional thermal history 
of the Universe. The answer is ``no" as simple estimates for the extra injected energy (due to the annihilation events with AQNs)
show. Indeed, the relative injection energy due to AQNs at temperature $T$ in comparison with average thermal energy 
$(T n_{\rm e})$ of the plasma can be estimated as follows, 
 \be
\label{E-injection}
\frac{1}{(T n_{\rm e})}\frac{dE}{dV}\sim    \frac{\left(R^2 T^2\right)\eta}{\alpha^2\la B\ra}\sim 10^{-19}\left(\frac{T}{1 ~\rm MeV} \right)^2.
\ee
see appendix in ref. \cite{Lawson:2018qkc}. The basic reason for this tiny rate is the same as 
discussed before: the cross section    is proportional to $R^2\sim B^{2/3}$, while  the number density of the nuggets is proportional to $\eta/ \la B \ra$ which results in a strong suppression rate (\ref{E-injection}). 
It is clear that such small amount of energy injected into the system will be quickly equilibrated within the system such the standard pre-BBN cosmology remains intact.  In other words, the conventional equation of state, and conventional evolution of the system is unaffected  by presence of AQNs.

\section{Post-BBN evolution}\label{low-T}

At temperatures below $T_\gamma \simeq m_e$ the electrons and positrons begin to annihilate causing 
their density to fall as $\sim e^{-m_e/T}$ until a major portion of the positrons get completely annihilated while the number densities for electrons $n_e$ and protons $n_B$ become  approximately equal  at $T_*\simeq 20$ keV. This is the consequence of the same ``charge separation" effect (replacing the ``baryogenesis" in AQN framework) when more antimatter than matter is hidden in form of dense nuggets  as reviewed in Section \ref{AQN}.  
 %At this point the  protons  replace the positrons as the dominant positive charge carriers in the plasma. 

 This regime when the AQNs are present in the plasma at $T_*\simeq 20$ keV has been recently discussed in \cite{Flambaum:2018ohm} in quite different context and for very different purposes.  To be more specific, it has been shown     that the primordial abundance of  Li and Be nuclei will be   depleted  in comparison  with    conventional BBN computations\footnote{The effect is  due to exponentially  strong enhancement of the capture probability (and subsequent  annihilation of Li and Be ions) by antinuggets. Technically the   effect  for heavy ions  with large $Z\geq 3$ occurs due to very strong enhancement factor $\sim \exp Z$, see  \cite{Flambaum:2018ohm} for the details.}. This   effect represents the resolution of the ``primordial Li puzzle" within AQN framework.   
 
 The main goal of the present work is very different, though the plasma regime surrounding the AQNs is the same with $T\lesssim T_* $.   In the present paper we study the survival pattern of the nuggets themselves, in contrast with 
the studies in \cite{Flambaum:2018ohm} when the main question was the analysis of  relative densities 
$\delta n_Z/n_Z$ of primordial nuclei with charge $Z$ as a result of the AQN presence in plasma.

We start our analysis by highlighting the basic features of the AQN electrosphere in the regime $T\lesssim T_* $
using  simple qualitative arguments. Later in the text we will support these arguments  by providing some analytical formulae. 
At $T\approx T_* $ when the external positron density essentially vanishes the boundary 
conditions for the AQNs' electrosphere fundamentally change resulting in a new charge distribution\footnote{One should  emphasize  that
the presence of electrosphere itself is a very generic phenomenon, and  its main features  
are determined by    the boundary conditions deep inside the nugget   where
the lepton's chemical potential is fixed as a result of the beta  equilibrium,  similar to 
 analysis   in  the context of strange stars, see \cite{Madsen:1998uh} for review.}. 
%We will assume that this transition occurs at a plasma temperature $T_*$ with the point at which the 
%positron and proton densities are equal suggesting $T_* \approx 20$keV.  
Below $T_*$ some 
fraction of the electrosphere positrons will be replaced by protons to fit with the new long distance, 
proton dominated boundary condition. The exact proton to positron ratio 
\exclude{$\chi \equiv  {N_p}/{N_{e+}}$} 
of the electrosphere will 
be determined by the rate at which captured protons are annihilated by the nuggets, positrons are annihilated by external electrons and the rate at which 
beta-processes can replace near surface positrons. 
\exclude{
It does not play any quantitative role in our discussions which follow
because we focus  on the physics  which occurs far away from the  surface. 
Therefore, we do not elaborate on this matter in the present work.
% In what follows we assume $\chi\sim 1$ to simplify our qualitative analysis. 
}

\exclude{
The  protons to positrons ratio  $\chi (T) $ in the electrosphere will be 
temperature dependent and its evolution will be determined by  the exact details of beta interactions 
within the quark matter and the rate of annihilation processes.
 The total number of   positrons expected to populate the electrosphere 
at $T =0$ may be estimated as $(\mu_0 R)^2$ where $\mu_0 = -e\Phi$ is the surface chemical 
potential which is typically at the tens of MeV scale for a wide range of quark matter models. This 
value is therefore the maximum number of protons which may be bound to the nugget at any given 
time. For typical nugget parameters we expect $(\mu_0 R)^2 \sim 10^{14}$. The majority of these 
positrons and protons are within a near surface region where the chemical potential is much larger than   
  $m_e$ and the positrons have a well defined fermi surface. These positrons are strongly 
electromagnetically coupled to the quark matter and are unlikely to be readily depleted even in a hot 
plasma background. 
}

Further from the quark surface the positrons are more weakly bound and thermal 
behaviour becomes important to the distribution. In this regime the density as a function of height 
scales as \cite{Forbes:2008uf,Forbes:2009wg},
\begin{equation}
\label{eq:Boltzmann}
n_{e^+} = \frac{T_N}{2\pi \alpha} \frac{1}{\left( z + \bar{z} \right)^2}, ~~~ 
\bar{z}^{-1} \approx m_e \sqrt{2\pi \alpha} \left( \frac{T_N}{m_e} \right)^{1/4}
\end{equation}
where the approximate value of $\bar{z}$ is taken from matching this solution to the numerical 
solution from higher densities.

\exclude{
If we assume that a substantial fraction of these thermally supported positrons are lost to the 
surrounding plasma we expect the AQNs to have a net electrical charge of approximately,
\begin{equation}
\label{eq:charge}
Q\sim 4\pi R_N^2 \int dz n(z) 
\sim 4\pi R_N^2 \frac{m_e T_N}{\sqrt{\pi \alpha}} \left( \frac{T_N}{m_e}\right)^{1/4} 
\end{equation}
which we will take as the approximate scale of the nugget charge though the actual $T$ dependance 
is likely more complicated and may depend on the dynamic properties of the system rather than 
simply the thermal average. 
}

The main  observation here is that  a $T\neq 0$ environment leads to ionization of the 
loosely bound positrons such that the antinuggets will be in a negative charged configuration  with  charge  $-Q$  estimated as follows
  \be
  \label{Q}
Q\simeq 4\pi R^2 \int^{\infty}_{z_0}  n(z)dz\sim \frac{4\pi R^2}{2\pi\alpha}\cdot \left(T\sqrt{2 m_e T}\right)~~~~~
  \ee
 where we assume that  some  loosely bound positrons   will be stripped off the electrosphere as a result of  non-zero temperature\footnote{We estimated the position of the cutoff  $z> z_0= (2 m_e T)^{-1/2}$ in  \cite{Flambaum:2018ohm}. 
 One should emphasize that all our estimates which follow are not very sensitive to the cutoff scale $z_0$.}. This negative charge of the antinugget implies that the protons from the plasma might be captured by the nugget by screening  the charge (\ref{Q}). It obviously implies that the effective cross section for capturing of the protons $\sim 4\pi R_{\rm cap}^2(T)$ will be drastically larger than $4\pi R^2$ from  our previous estimates (\ref{eq:hot_rate}), (\ref{eq:hot_col}) when the electrosphere is entirely made of the positrons, not protons.

In principle the distribution of protons surrounding the nugget should be determined through 
a Thomas-Fermi computation similar to that performed in \cite{Forbes:2009wg} but allowing 
for the presence of protons as well as positrons and using the early universe plasma density 
as the $r\rightarrow \infty$ boundary condition. However, for present order of magnitude estimates  we will assume 
a simple power law scaling   with exponent $p$,
\begin{equation}
\label{eq:Pdens}
n_p(r) = n_0 \left( \frac{R}{r} \right)^p
\end{equation}
with  the normalization $n_0$ set to match the total charge given in expression (\ref{Q}).     This assumption is consistent with our numerical studies \cite{Forbes:2009wg} of the electrosphere with $p\simeq 6$ for positrons. It is also consistent with conventional Thomas-Fermi model at $T=0$, see  \cite{Flambaum:2018ohm} for references and details. 
     We keep parameter $p$ to be arbitrary to demonstrate  that our main claim is not very sensitive to our assumption on numerical value of $p$. With these assumptions the baryon number density $n_0$ in close vicinity of the nugget can be estimated as follows,
\begin{equation}
n_0 = \frac{p-3}{4\pi R^3} Q.
\end{equation}
One should note that  the behaviour of the proton cloud may deviate significantly 
from expression (\ref{eq:Pdens}) at very small and very large radii, however we simply want to 
determine the approximate scale over which electromagnetic effects will act. In this context the 
simple form of expression (\ref{eq:Pdens}) should be sufficient. 
%In fact, the power law scaling will 
%be further suppressed at large distances so this should be taken as an upper limit of the effective 
%interaction range. 
This behaviour will continue until the proton density matches that of the 
surrounding plasma which gives us a radius for the over density of protons surrounding by the nugget.
\be 
\label{eq:R_cap}
\left(\frac{R_{\rm cap}}{R}\right)^p \sim \left( \frac{n_0}{\eta n_{\gamma}} \right)  \sim 10^{12}\cdot \left(\frac{20~ {\rm keV}}{T}\right)^{3/2}.
\ee
This increase in the effective scattering length of the nuggets will boost the number of interactions 
and may result in an increased annihilation rate. Most importantly these captured protons will 
spend an extended amount of time near the surface of the AQN giving them an increased 
opportunity to annihilate.   Again, we stress that this increased rate of proton capture is effective only 
after the positrons are fully annihilated and the protons are the only positive charge carriers in the 
plasma which happens at $T\simeq T_*\simeq 20$ keV.

In particular for $p\simeq 6$   the   effective capture distance 
$R_{\rm cap}$ is of order
\be
\label{R_cap2}
   R_{\rm cap} \simeq    10^2\cdot \left(\frac{20~ \rm  keV }{T}\right)^{\frac{3}{2p}} R ,   
\ee
which of course drastically changes the collision rate as will be estimated below. The scaling (\ref{R_cap2}) holds as long as the thermal equilibrium between the nuggets and surrounding plasma is maintained. Formula  (\ref{R_cap2}) breaks down 
at sufficiently large $R_{\rm cap}$  when the power law scaling  (\ref{eq:Pdens})  is replaced by 
the  exponential behaviour due to the Debye screening. Numerically, 
the Debye screening becomes operational at $R\simeq 10~ R_{\rm cap}$ at $T\simeq 20$ keV, see Appendix A in \cite{Flambaum:2018ohm}.

We may now perform the same estimates  as in expressions (\ref{eq:hot_rate}) but using the 
larger capture cross section of equation (\ref{R_cap2}), i.e. 
\begin{equation}
\label{eq:cool_rate}
\begin{aligned} 
\Gamma_{\rm col}(T) = 4\pi R_{\rm cap}^2(T)  n_B (T) v_B(T). 
%\approx 4\pi R_N^2 \left(\eta n_{\gamma}\right)^{1-1/p} n_0^{1/p} \sqrt{\frac{2T}{m_p}} c 
\end{aligned}
\end{equation}
The total number of collisions during this time is saturated by the highest temperature $T\simeq T^*$ such that 
the integral can be estimated as follows
\be
\label{eq:cool_col}
&& N_{\rm col}(T) = \int_0^{T_*} dT~ \frac{dt}{dT} \Gamma_{\rm col}(T) \\
  &\sim& 10^{24} \left(\frac{T}{20 ~\rm{keV}}\right)^{(\frac{3}{2}-\frac{3}{p})} \left(\frac{R}{10^{-5}{\rm{cm}}}\right)^{(2-\frac{2}{p})}  \nonumber
\ee
where we use $p=6$ for numerical estimates. The expression   (\ref{eq:cool_col})  represents a full  analog of the estimate (\ref{eq:hot_col}) obtained for pre-BBN epoch. 

\exclude{
where we define the integral,
\begin{equation}
F_p(T_*) = \int_0^{T_*} \frac{dT}{T_*} \sqrt{\frac{T}{T_*}} 
\left(\frac{T_N}{T_*}\right)^{5/2p} \left(\frac{T_*}{T}\right)^{6/p}
\end{equation}
which holds most of the uncertainty in the long range behaviour of the proton distribution 
and the relation between the temperature near the core of the nugget and that in the plasma. 
We will assume that $p>3$ and $T_{\gamma}, T_N \lesssim T_*$ so that the exact 
value of this integral is an order one correction factor within the margin of error of these 
calculations.

If we consider the case where $p=6$ expression \ref{eq:cool_rate} is approximately,
\begin{equation}
\label{eq:cool_col}
N_{col} \approx 4\times 10^{24} \left(\frac{T_*}{25{\rm{keV}}}\right)^{3/2} 
\left( \frac{R_N}{10^{-5}{\rm{cm}}} \right)^{5/3}
\end{equation}
}
While the number of scatterings occurring in this low temperature regime is slightly  
below the number occurring just after nugget formation as estimated in equation (\ref{eq:hot_rate}) we 
expect the evolution of the AQN baryon number to be dominated by these low energy collisions in 
which the AQN and baryonic matter temporarily form a bound state. This is because, as argued above, 
collisions at tens of MeV are highly likely to result in elastic scattering while electromagnetically bound 
protons have a much larger opportunity to come into overlap with the quark modes of the colour 
superconductor and eventually annihilate. 

The similarity in the total number of collisions  in the  estimates (\ref{eq:hot_col}) and 
(\ref{eq:cool_col}) at 
$T_{\rm form}\simeq 40~{\rm MeV} $  and $T_*\simeq 20$ keV correspondingly can be easily 
understood from the following simple observations. 
The baryon number density in plasma scales as $T^3$, the proton's velocity in plasma scales as $T^{1/2}$, the cosmic time scale as $T^{-2}$. All these factors result in drastic changes in the rate between $T_{\rm form}\simeq 40$ MeV and $T_*\simeq 20$ keV with approximate suppression factor:    $(T_*/T_{\rm form})^{3/2}\sim 10^{-5}$.   
However, these substantial  suppression  in the temperature is  mostly  compensated by enhancement  in effective cross section $(R_{\rm cap}/R)^2\sim 10^{4}$, see estimate (\ref{R_cap2}). 
These two  effects work in opposite directions  which explains our estimates for  the collision rates (\ref{eq:cool_col})  and   (\ref{eq:hot_col}) being  numerically close.

We summarize  this section with the following comment. The  total number of collisions   
(\ref{eq:cool_col}) 
is still much smaller than 
a typical baryon charge $\la B\ra \sim 10^{25}$ of the nuggets, such that the majority of the nuggets will survive the post-BBN epoch  as  only small portion of the collisions will eventually lead to the annihilation events. Therefore, from these estimates we conclude 
 that the post-BBN  epoch does not modify the primordial spectrum of the AQNs.

 At this point a thoughtful  and careful reader may wonder how it could happen that the number of annihilation events 
 estimated above is sufficiently small  that  all nuggets with $B> 10^{24}$ can easily survive  the 
 unfriendly hot and dense environment of early universe 
according to estimates (\ref{eq:cool_col})  and   (\ref{eq:hot_col}). At the same time, it has been argued recently  in   \cite{Zhitnitsky:2017rop, Zhitnitsky:2018mav, Raza:2018gpb} that the nuggets of all sizes will experience complete annihilation  in the solar corona when the AQNs enter the solar atmosphere. How  these two claims could be consistent? 
 We refer the readers to  Appendix \ref{turbulence} specifically  addressing    this question where we  emphasize a number of crucial differences between the two cases.  
 
 The only comment we would like to make here is that the drastic enhancement of  the rate of annihilation 
 in solar corona is due to the propagation of the AQN with supersonic speed (above the escape velocity $v> 600$ km/s at the solar  surface) in the ionized plasma with a very large Mach number $M=v/c_s\simeq 10$, where $c_s$ is the speed of sound in the solar atmosphere. 
 It is well known that a moving body with such a large Mach number will inevitably generate a 
 shock wave and accompanying it a temperature discontinuity with turbulence in vicinity of a moving body.  As a result of this complicated non- equilibrium  dynamics    the effective cross section may be drastically increased  in the course of shock wave propagation due to the capture (with subsequent annihilation) of a large number of ions from plasma. These features  of the AQNs  in the solar corona should  be contrasted with AQN's  adiabatic evolution in the plasma of the early  Universe with  its relatively slow  evolution.   
 
 \section{Post-recombination evolution}\label{recombination}
 The integral (\ref{eq:cool_col}) is saturated by the highest possible value $T\simeq T_*\simeq 20$ keV because the largest collision rate 
occurs precisely at that time. As the temperature slowly decreases due to the Universe's expansion  one should not expect any dramatic changes until the recombination epoch at $T\simeq 0.3$ eV.
At this point  the universe becomes  neutral and the scattering cross section of matter with 
the AQN no longer receives a boost from electromagnetic effects analogous to (\ref{R_cap2}).  Therefore, these generic arguments suggest that the collision rate diminishes even faster after recombination impling that  the size distribution of the AQN essentially does not change during that epoch. These generic arguments obviously do not apply to violent environments during galaxy or star formation (which occur during this epoch) and which will be analyzed later in this section. 

In spite of this relatively dilute environment, the rare events of annihilation of the AQNs with surrounding baryons still occur even at such low density. The corresponding radiation due to the annihilation processes, while negligible in comparison with the dominating CMB radiation, may nevertheless leave some imprints  which could be observed today, as argued in \cite{Lawson:2018qkc}. This is due to some specific features of the spectrum characterizing the AQN annihilation events: 
the low energy tail of the radiation due to the annihilation processes with the nuggets has spectrum $\sim \ln \nu$ which should  be contrasted with conventional CMB black body radiation characterized by $\nu^2$ behaviour at low $\nu\ll T$, see  \cite{Lawson:2018qkc} for the details. 
 
 As we mentioned above,  the violent environment during the galaxy or star formation  requires a special treatment  because it may potentially  change  the generic argument that this epoch is essentially irrelevant to the AQN's survival pattern. 
In what follows we compare the environment during the galaxy and  star formation with corresponding features in the solar corona (where it has been argued that complete annihilation of the AQNs occurs)   in the context of the AQN annihilation rate. The outcome of this comparison and  corresponding  conclusion will be formulated at the very end of this section. 
 
 The complete annihilation of the AQNs in the solar corona as discussed in  \cite{Zhitnitsky:2017rop,Raza:2018gpb} and reviewed in Appendix \ref{turbulence} is the direct consequence  of a few  factors:\\
 1. relatively large density in the transition region $n\sim 10^{11} {\rm cm^{-3}}$;\\
 2. very large velocity of the nuggets  on the solar surface, $v > v_{\odot} = \sqrt{2GM_{\odot}/R_{\odot}}\simeq 600 $ km/s. This is  of course  a result of  strong  gravitational forces $\sim M_{\odot}$  localized on a relatively small distance $R_{\odot}$;\\ 
  3.   large velocity $v$  greatly exceeds the speed of sound $c_s\simeq \sqrt{T/m_p}$.   It implies that  the Mach number is very large, $M\equiv v/c_s\gg 1$ such that shock waves inevitably form;\\
 4. high ionization of the plasma    due to large temperature $T\simeq 10^6$K in the transition region.
   
 The combination of these   factors   lead to complete annihilation of the AQNs  as reviewed in 
  Appendix \ref{turbulence}.   While individual (violent) condition from the list above  may emerge during     the galaxy  or star formation, the combination of all 4  elements  does not occur, in general,  during this epoch. Therefore, we do not expect any considerable modification of the size distribution of the nuggets  after the recombination.
  
  Indeed, the baryon density during the structure formation epoch does not exceed $n_B\sim 1~{\rm cm^{-3}}$. Furthermore, 
  the typical velocities of particles in the  gas are the same order of magnitude as the speed of sound, i.e. $v\sim c_s\sim  10^2 ~{\rm km/s}$ such that one should use a conventional formula  for estimation of the collision rate   without any additional enhancement factors related to Mach number $M$, i.e.
 \be
 \label{galaxy-1}
 \nonumber
 \Gamma_{\rm col} \sim 4\pi R^2  n_B  v_B \sim 10^{-2}s^{-1} (\frac{n}{{\rm 1~cm^{-3}}}) (\frac{v}{100{\rm km/s}} ). 
\ee
The total number of collisions $N_{\rm col}$  during the Hubble time $H^{-1}$ at redshift $z\sim 10$ is of order 
\be
 \label{galaxy-2}
  N_{\rm col}\sim \Gamma_{\rm col}\cdot H^{-1} \sim 10^{14} \left(\frac{n}{{\rm 1~cm^{-3}}}\right) \left(\frac{v}{100~{\rm km/s}} \right), ~~~
\ee
which represents a tiny portion of the average baryon charge $\la B\ra \sim 10^{25}$   of a nugget. Furthermore, even if in some small regions the relative velocities of the AQNs and   baryons exceed the speed of sound $c_s$ it does not lead to shock wave formation  similar to our discussions in the solar corona (reviewed in Appendix \ref{turbulence}). This is  because the shock wave phenomenon is  
based on  effective description of the system when the  hydrodynamical description 
  is justified, which implies that a typical distance between the particles $\sim n^{-1/3}$
must be much smaller than the size of a moving body $\sim R$. This approximation   is obviously  badly violated   for the AQNs during the structure formation epoch. Therefore,  according to our estimate   $N_{\rm col} \ll \la B\ra \sim 10^{25}$, and we conclude that the size distribution of the AQNs is not modified during the era of structure formation. 

A similar conclusion also holds for  another violent epoch (star formation)  which could  also potentially modify the AQN size distribution.  The corresponding analysis can be separated  in two different stages: the final stage of formation when typical  parameters are similar to our analysis of the Sun, and the initial  stage of star formation characterized by the density ranging
from $n\sim 10^{15} {\rm cm^{-3}}$ to $n\sim 10^{0} {\rm cm^{-3}}$ depending on size of infall cloud ranging from 
$r\sim 10^{-1} {\rm AU}$ to $r\sim 10^{6} {\rm AU}$, see \cite{Yoshida:2006bz}.

We start our estimates from the final stage of   formation when the stars assume their final form.
In this case all the nuggets which will be captured by a star will be completely annihilated similar to our studies  of the sun.
However, the portion of the AQNs which will be captured by the stars is very tiny in comparison with total number of nuggets. The corresponding portion of the affected nuggets  can be estimated in the terms of the capture impact parameter $b_{\rm cap}$ which is typically only few times the star's size $R_{\star}$. 
\be
  \label{capture_star}
  b_{\rm cap}\simeq R_{\star}\sqrt{1+\gamma_{\star}}, ~~~~ \gamma_{\star}\equiv \frac{2GM_{\star}}{R_{\star}v^2},
  \ee
where $v\sim 10^{-3}c$ is a typical velocity of the nuggets  far away from the star.
The rate of the total    mass  annihilation  $dM_{\rm ann}/dt$ of all nuggets captured (and consequently annihilated) by the star    can be estimated as follows
 \be
  \label{total_power_mass}
  \frac{dM_{\rm ann}}{dt}\sim 4\pi b^2_{\rm cap}  v  \rho_{\rm DM}   
  \simeq 3 \cdot10^{30} \left(\frac{v}{10^{-3}c}\right) \frac{m_p}{\rm  s}, 
    \ee
where we used the solar parameters for the numerical estimates and assumed that $\rho_{\rm DM}$ is saturated by the AQNs\footnote{One should comment here that in case of the Sun the energy released as a result of annihilation events with the rate 
 (\ref{total_power_mass}) represents   approximately $10^{-7}$ of the total solar luminosity radiated from solar corona in the  form of the EUV and x-rays, which represents the resolution of the solar corona  heating problem within AQN scenario as  suggested in \cite{Zhitnitsky:2017rop,Raza:2018gpb}.
 }. The upper limit of total mass annihilated by a single  star can be estimated by multiplying (\ref{total_power_mass}) to the
total life time of  star which can be approximated as $H^{-1}$, i.e.
\be
\label{final_star}
M_{\rm ann}\leq  \frac{dM_{\rm ann}}{dt}\cdot H^{-1}\sim 10^{48} m_p\sim 10^{21}{\rm kg}. 
\ee
 Estimate (\ref{final_star}) should be compared with total mass of the star $M_{\star}\sim M_{\odot}\sim 10^{30}{\rm kg}$.
  As the stars represent only a fraction of the total baryonic matter of the Universe, and  the DM is 5 times the total baryonic matter, one can infer   from   (\ref{final_star})  that  the annihilated portion of DM (due to the capturing by stars)  represents only a small portion  $\lesssim  10^{-10}$    of the total   dark matter material of the Universe.

 We now turn to the estimates of the AQN annihilation pattern   during the initial  stage of star formation. In this case the DM nuggets passing  through   infall cloud experience the annihilation events. The corresponding total    annihilated baryon charge  for a single nugget  (as a result of this passage)  can be estimated as follows, 
  \be
 \label{star-1}
  N_{\rm col} \sim \pi R^2  n_B  L\sim   10^9\cdot (\frac{n_B}{{\rm 1~cm^{-3}}}) \cdot (\frac{L}{10^6 AU}), 
\ee
which represents a tiny portion of the average baryon charge $\la B\ra\sim 10^{25}$. In estimate (\ref{star-1})
   we used the most generic configurations when AQNs enter a large region with $L\sim 10^6$AU  characterized by  $n_B\sim {\rm 1~cm^{-3}}$, while passage of the nuggets through small well-localized  region $L\sim 10^{-1}$ AU with high density $n_B\sim {\rm 10^{15}~cm^{-3}}$ is highly unlikely as it represents a very small portion of the total AQN flux. 
   But even in this case     the total number of annihilation events $N_{\rm col} \sim 10^{17}$  remains small in comparison with average baryon charge $\la B\ra\sim 10^{25}$.
   
   We conclude this section with the following comment: The violent events such as galaxy formation or star formation after recombination do not drastically modify the size distribution of the nuggets, similar to our previous analysis devoted to the different epochs in the Universe evolution. The basic reason for this conclusion is that the nuggets which can experience compete annihilation represent only small portion of the entire population according to (\ref{final_star}), while the majority of the nuggets will loose a very tiny portion of their baryon charge during the Hubble time according to (\ref{galaxy-2}).

\section{Present Day Mass distribution}\label{present}
After recombination the size distribution of the AQN to be essentially fixed 
  until the present day\footnote{Obviously the distribution may develop 
localized anisotropy in regions of particularly high matter density such as stars or planets, but these local effects are 
not the subject of the present work.}. 
We may formulate the change in baryon 
number as a result of post-formation annihilation as,
\begin{equation}
\label{eq:total_dB}
\Delta B = f_1 N_{\rm col}(T>T_*)  +f_2 N_{\rm col}(T<T_*) 
\end{equation}
where $f_1$ and $f_2$ are the fractions of collisions in the positron dominated and 
proton dominated phases which result in the annihilation of a unit of baryon charge from the 
AQN with the collision rates taken from expressions (\ref{eq:hot_col}) and (\ref{eq:cool_col}) correspondingly.
If we assume an original nugget distribution defined by $B_{\rm min}$, $B_{\rm cut}$ and 
$\alpha$ as discussed in section 
\ref{subsec:plotsNB} we may use expression (\ref{eq:total_dB}) to translate the distribution formed 
above $T\approx 40$ MeV to a present day mass distribution. Note that, since $R \propto B^{1/3}$
the first term in expression (\ref{eq:total_dB}) scales as $B^{2/3}$ according to (\ref{eq:hot_col}), while the second scales as 
$B^{5/9}$ according to (\ref{eq:cool_col}). 

We may then write the late time baryon number distribution based on result (\ref{eq:slope_prop}) 
obtained from percolation theory as follows\footnote{\label{footnote:B_min}Formula (\ref{eq:slope_prop}) formally holds only for asymptotically large 
$B\gg B_{\rm min}$, but in fact remains valid for almost entire region of $B$ with the exception of a small  region in  vicinity of $B_{\rm min}$, see Fig. \ref{fig:alpha} in Appendix \ref{appen:NB}.},
\begin{equation}
\label{eq:distribution}
\frac{dN}{dB} = N_0 \left( \frac{B_{\rm min}}{B}\right)^{\alpha}, ~~
B\gg B_{\rm min}, 
\end{equation}
where $B$ is the present day baryon number of the AQN and we have ignored the small 
portion $\Delta B$ related to the annihilation processes as discussed above. 

\exclude{ 
$B<B_{\rm min}-\Delta B$ will be fully annihilated. 
Note that while this is expression is only the appropriate when $dR/R = \frac{1}{3}dB/B$ 
is small any scenario in which $\Delta B$ is large relative to $B_{min}$ will shift the peak of the 
baryon distribution down to zero which is observationally disallowed.

The fact that only the antiquark 
AQN undergo annihilations after formation also implies that the initial baryon number difference 
between the nugget and antinugget distributions may differ by amount  $\Delta B$
representing the annihilating portion of the baryon charge as defined in (\ref{eq:total_dB}).
}

For a given environment the constant $N_0$ may be fixed if the AQNs are assumed to provide 
the entire dark matter mass\footnote{As we reviewed  in Section \ref{AQN}  the conventional axion production due to the misalignment mechanism
and domain wall decay always accompany the nuggets' formation and contribute to total DM density, see Fig.\ref{phase_diagram}. However, the relative 
portion   of these well studied mechanisms  to $\rho_{\rm DM}$ strongly depends on the axion mass $m_a$ and it is likely to be negligible for  sufficiently large axion mass, see \cite{Ge:2017idw} with corresponding plots.}. We also assume that one and the same $\alpha$ describes the mass distribution for all values of $B$.
It may or may not be a correct assumption as some nanoflare's models fit  the solar corona heating  with different exponents $\alpha$ for small and large values of $B$, see below with additional comments. 

With the assumptions just formulated  one can represent $\rho_{\rm DM} $   as follows,  
\begin{equation}
\label{rho_DM}
\rho_{\rm DM} = \int^{B_{\rm cut}}_{B_{\rm min}}  m_p B \frac{dN}{dB}dB  
\end{equation}
where ${dN}$ has a physical meaning of the AQN number density per unit volume\footnote{Note that the definitions of $N$ and $N_{0}$ here are slightly different from that defined in section~\ref{sec:distribution} where $N$ and $N_{0}$ are the total numbers there rather than number densities.} per baryon charge  interval $dB$, while  $m_p B ({dN}/{dB})$ has physical meaning of the mass density per unit volume hidden in form of the nuggets with baryon charge from $B$ to $B+dB$. The relation (\ref{rho_DM})
  allows  us to solve for the normalization factor in expression (\ref{eq:distribution}) for different exponents $\alpha$'s:
\be
\label{N_0}
N_0 &=& ( \alpha-2)\frac{\rho_{\rm DM}}{m_p B_{\rm min}^{2}} ~~~~~~~~~~~~~~~\alpha> 2  \\
N_0 &=& \frac{\rho_{\rm DM}}{m_p B_{\rm min}^2 {\rm{ln}}(B_{\rm cut}/B_{\rm min})}~~~~~~\alpha = 2 \nonumber \\
N_0 &=& (2-\alpha) \frac{\rho_{\rm DM}}{m_p B^2_{\rm cut}} \left(\frac{B_{\rm cut}}{B_{\rm min}}\right)^{\alpha} 
~~ \alpha< 2  \nonumber
\ee
where we have again assumed that $\Delta B\ll B$ represents a small fraction of the initial baryon
number as argued to be the case in previous sections \ref{ssec:PreBBN} and  \ref{low-T}. 

Note that $\alpha = 2$ marks the slope at which the distribution transitions from being 
mass dominated by the bottom end to the higher end. Distributions with $\alpha > 2$ are largely 
defined by the mass scale $B_{\rm min}$ while the typical mass scale for shallower sloped distributions 
with $\alpha<2$ is dominated by $B_{\rm cut}$.

\exclude{This also lets us estimate the number density of AQN with a size above $B_*$ as,
\begin{equation}
\label{eq:heavy_frac}
n(B>B_*) = \frac{2+\alpha}{1+\alpha} \frac{\rho_{DM}}{m_p B_{min}} \left( \frac{B_*}{B_{min}}
\right)^{1+\alpha}
\end{equation}
for any $B>B_{min}$.}

We may now ask, given the profile of expression (\ref{eq:distribution}) what range of parameters 
are consistent with observational constraints discussed in Section \ref{sec:Constraints}? As argued above we expect 
that $f_1 << f_2$ so that the majority of annihilations occur below $T\sim20$ keV, if this is the case 
then the parameter space of the AQN model may largely be defined by four variables; $B_{\rm min}$, 
$B_{\rm cut}$, $\alpha$ and $\Delta B \sim f_2$. As discussed in section \ref{subsec:plotsNB} 
these parameters are not well 
constrained from the theoretical end. Indeed, while the theoretical analysis predicts a generic  
power law behaviour (\ref{eq:slope_prop})  the 
numerical value for the exponent $\alpha$ is expressed in terms of parameter $\delta$ according to eq.  (\ref{eq:alpha_nugget}) 
which itself describes some features  of the bubble's formation during the QCD transition  at 
$T\sim T_c$,   which are basically unknown  in strongly coupled QCD.

\exclude{In principle $B_{min}$ is constrained from above by the 
baryon content of the coherent axion field at the time of formation so that 
$B_{min}<\Lambda_{QCD}^3m_a^{-3}$ which, for the allowed axion mass window could range 
from $10^{30}$ to $10^{42}$, however this value certainly represents a large over estimate of 
the baryon capture efficiency. }

From the observational end the constraints are less trivial and much more interesting. 
First of all, there are constraints on $\alpha$ which come 
primarily from solar data.  As we  
discussed in section \ref{ssec:Solar}, the solar corona measurements are sensitive to the full distribution 
rather than simply the average value $\la B \ra$. 
If the nuggets are to offer an explanation to the solar heating problem as argued 
in \cite{Zhitnitsky:2017rop, Zhitnitsky:2018mav, Raza:2018gpb} then we require that the majority 
of heat input comes from lower energy unobservable events and thus must have $\alpha>2$.
This option is consistent with analysis of the nanoflare distribution performed in ref.\cite{Pauluhn:2006ut} where the authors claim that the best fit to the data is achieved with $\alpha\simeq 2.5$
while numerous attempts to reproduce the data with $\alpha<2$ were unsuccessful. 

Another option advocated in ref. \cite{Bingert:2013}  is that the nanoflare distribution consists two different exponents:
The events below $E\leq 10^{24} {\rm erg}$ are described by $\alpha\simeq 1.2$ while the higher energy events  
with $E\geq 10^{24} {\rm erg}$ are described by $\alpha\simeq 2.5$. In terms of baryon charge distribution (\ref{eq:distribution}) the
model   \cite{Bingert:2013}  corresponds to the nugget mass distribution with $\alpha\simeq 1.2$ and  cutoff scale near $B_{\rm cut}\sim10^{27}$.
The higher mass nuggets with $B_{\rm cut}\geq 10^{27}$ are described by   $\alpha\simeq 2.5$.  
 
 \exclude{
In the weaker case where the AQN do not fully provide the coronal heating there is still a limit 
on the number of observable flare-like events with energies above $\sim 3\times 10^{24}$ergs.
This gives a constraint on the total number on of AQN impacts on the sun involving a baryon
number greater than $B\sim 10^{27}$. The rate of these events over the entire surface of the sun 
is estimated to be $\sim 300$s$^{-1}$ \cite{Benz:1998} though the vast majority of these are 
associated with known solar activity and not a direct result of the impact of a quark nugget.
The rate of flares due to the impact of 
an AQN with a baryon number above $B>10^{27}$ is $n(B>10^{27}) v_g$ with $n$
given by the integral of expression (\ref{eq:distribution}) from $B=10^{27}$ up to the cutoff scale 
and $v_{\rm AQN} \approx 200$ km/s being a  typical DM  particle velocity. The nano-flare models discussed above and advocated in refs \cite{Pauluhn:2006ut} and   \cite{Bingert:2013} automatically satisfy this constraint. The baryon charge distribution (\ref{eq:distribution})
in the AQN framework (reproducing the corresponding nano-flare distributions \cite{Pauluhn:2006ut} and   \cite{Bingert:2013} as discussed above)  also automatically satisfies this constraint. 
}

Assuming that the distribution (\ref{eq:distribution}) can be approximately used in close vicinity of $B_{\rm min}$
one can relate  the  average baryon number $\la B \ra$ and $B_{\rm min}$ as follows  
\begin{equation}
\label{B_min}
\la B \ra \approx \frac{\alpha-1}{\alpha-2} B_{\rm min}, ~~~ \alpha>2 
\end{equation}
so that, for any $\alpha>2$ the average  $\la B \ra$ and $B_{\rm min}$ are at the same scale.
One should not literally use the relation (\ref{B_min}) as  the distribution (\ref{eq:distribution}) cannot be numerically trusted in close vicinity
of $B_{\rm min}$, see footnote \ref{footnote:B_min} with comments. If one uses the observational constraint on $\la B\ra$ from (\ref{direct})
one can impose the constraint on  $B_{\rm min}\gtrsim 10^{24}$ from (\ref{B_min}).

\exclude{
  In the case where $\alpha<2$ the 
mass distribution tilts towards the upper end of the distribution and the value of $B_{\rm min}$ is 
not well constrained by astrophysical observations or direct detection. 
However,  populations with $\alpha<2$ could be  in tension with the solar observations unless the 
distribution is cutoff below the scale at which a nugget annihilations would become detectable as explicitly 
realized in nano-flare model  \cite{Bingert:2013} with two different exponents describing the low and high energy ends of the distribution
with a position of a knee around  $B\simeq 10^{27}$. 
}

For the AQN population to survive as high mass dark matter 
candidates we require $B_{\rm min}>\Delta B$ which, from expression (\ref{eq:cool_col}), suggests a 
lower bound below $\sim 10^{24}$ though, as argued above this scale 
%may be strongly effected 
%by uncertainties in the long range electromagnetic effects. In particular expression \ref{eq:cool_col} 
considers the number of  collisions between background plasma with  
 an AQN rather than the number of annihilation events $\Delta B$. The only robust constraint on 
 $B_{\rm min}$  comes from the observational side (\ref{direct}) which can be expressed in terms of the $\la B\ra$  as discussed above.

These considerations lead us to two general classes of AQN distributions which will be consistent 
with all known direct and indirect constraints. These models are essentially  
equivalent to a   variety of the  nano-flare distributions \cite{Pauluhn:2006ut,Bingert:2013} expressed in terms of the 
baryon charge number with the only additional condition. The nano-flare models must satisfy   condition $B_{\rm min}>10^{24}$ to be consistent with  independent observational constraint (\ref{direct}). There are plenty of models among   \cite{Pauluhn:2006ut,Bingert:2013}  which satisfy this condition and saturate the required energy budget for the corona heating,  and we shall not elaborate here on this topic. 

We consider that this  phenomenon when the allowed window in the baryon charge $B$ and the fitted energy spectrum for nanoflares are overlap with identification $E\simeq  2m_pc^2 B$   as a  highly nontrivial self-consistency check of the AQN framework. On one side this window  represents   a cumulative constraint  from  a number of astrophysical, cosmological, satellite  and ground based observations and experiments  reviewed in section \ref{AQN}  which are consistent with analytical results  based on percolation theory discussed  in sections  \ref{sec:RTE} and \ref{sec:distribution}.
On other hand, this window  largely overlaps    with the  constraints originated from completely independent  physics:    the solar corona heating  when nanoflare distribution (\ref{distribution})  with energy $E$ is identified with AQN distribution  (\ref{eq:distribution}) with the baryon charge $B$.

\section{Conclusion}
The main results of this work can be formulated as follows:\\
1. We used  an  approach coined as 
\textit{envelope-following} method to overcome a common numerical  problem with drastic separation of scales in the  system.
In our case the scales are: the QCD scale $\sim\Lambda_{\rm QCD}$, the axion scale $\sim m_a$ and the cosmological time scale $t_0\sim 10^{-4} {\rm s}$.
 The results support our original assumptions that   the chemical potential inside the nugget indeed assumes a sufficiently large value  $\mu_{\rm form}\gtrsim 450$~MeV during this long cosmological evolution. This magnitude  is consistent with  formation of a 
  CS phase, see Fig.\ref{fig:RTE};\\
  2. The nuggets complete their formation precisely in the region of $T_{\rm form}\approx 40 $ MeV where they should, see Fig.\ref{fig:RTE},  as it corresponds to the temperature where the baryon to photon ratio $\eta$ assumes its present value (\ref{eta});\\
  3. Items 1 and 2  represent a highly nontrivial   consistency check of the AQN framework when three drastically different scales ($\Lambda_{\rm QCD}$, axion mass $m_a$ and cosmological time scale $t_0$)      ``conspire" to produce a self-consistent picture;   \\
  4. We argued that  the nugget's distribution must have the algebraic behaviour  (\ref{eq:slope_prop}) as a direct consequence of a generic  feature of the percolation theory. The exponent $\alpha$ cannot be predicted theoretically, but can be expressed, according to (\ref{eq:alpha_nugget}), in terms of another parameter which is sensitive to the axion domain wall  formation during the QCD epoch;\\
  5. We argued that the nuggets   survive both: the pre-BBN and post-BBN evolutions (as estimations  in sections \ref{ssec:PreBBN} 
  and \ref{low-T} show) as long as they sufficiently large to  satisfy the observation constraint (\ref{direct}). The essential reason for this is that the fraction of an AQN annihilated scales with the 
cross section to mass ratio $\sim B^{-1/3}$ and thus nuggets of sufficiently large $B$ will 
remain largely unaltered by annihilation events after their formation;\\
  6. We argued that the present  day baryon charge distribution (\ref{eq:distribution}) is consistent with nanoflare distribution (\ref{distribution}) which was fitted to describe the solar corona observations. This    represents a highly nontrivial consistency check of the proposal  \cite{Zhitnitsky:2017rop, Zhitnitsky:2018mav, Raza:2018gpb} that the AQNs made of antimatter are the nanoflares postulated long ago.   
  
It is the central claim of this work that there exists a larger amount 
of the allowed parameter space across which the AQN model is consistent with all available cosmological, astrophysical, satellite and ground based constraints. While the model was invented long ago to    explain  the observed relation  $\Omega_{\rm dark}\sim \Omega_{\rm visible}$, it may also   explain a number of other observed phenomena, such as the  excesses of galactic emission in different frequency bands
as reviewed in section \ref{AQN}. This model also offers a resolution of the so-called ``Primordial Lithium Puzzle" 
and  the 70 yers old ``The Solar Corona Mystery" as mentioned in Introduction when one uses precisely the {\it same parameters}
and the same distribution  (\ref{eq:distribution}) advocated in the present work. However, all these manifestations of the AQN model are indirect in their nature. Therefore,    one can always find tons of alternative  explanations for the same phenomena.

\exclude{
In this paper we have produced an estimate of the initial distribution of masses (or 
equivalently baryon charges) associated with the nuggets of the AQN dark matter model. 
While the general form of this distribution is rather complicated it simplifies to a 
power law under a reasonable set of approximations allowing us to describe the 
initial distribution with only three parameters ($B_{min}, B_{cut}$ and $\alpha$.) Furthermore 
we have demonstrated that for a large and physically interesting range of these parameters 
the AQN are capable of surviving in the hot and dense conditions of the early universe.  

Our analysis represents the first prediction of the full mass spectrum of AQN dark matter 
rather than simply its average value. This has allowed us to revisit the known observational 
constraints on high  mass dark matter and reformulate them in terms of the parameters 
of the distribution function.
}

In this respect the recent  proposal advocated in \cite{Fischer:2018niu,Liang:2018ecs,Lawson:2019cvy} to search for  the axions (which will be  inevitably produced as a result of   the annihilation processes of the antimatter nuggets with surrounding matter) is the direct 
manifestation of the AQN model. 
These axions will be emitted when the AQNs get disintegrated in the solar corona. The axions will be also emitted when the nuggets hit the Earth and continue to propagate deep underground  and loosing the baryon charge  accompanied by emission of   the axions.    
In fact, the observation of these axions with very distinct spectral properties in comparison with conventional galactic axions 
 will be the smoking gun supporting the entire AQN framework.     We finish this work on this positive and optimistic note.

\section*{Acknowledgements} 
 This work      was supported in part by the Natural Sciences and Engineering 
Research Council of Canada.

\appendix
\section {Nugget evolution from $T_{0}$ to $T_{\rm form}$}\label{appen:evolution}
In this section, we are going to discuss the time evolution of a nugget from $T_{0}$ to $T_{\rm form}$ both analytically and numerically.  We start with the Lagrangian that dominates the nugget evolution
\begin{equation}\label{eq:lagrangian}
\mathcal{L}=\frac{4\pi\sigma_{\rm eff} R^2}{2}\dot{R}^2-4\pi\sigma_{\rm eff} R^2+\frac{4\pi R^3}{3}
\Delta P.
\end{equation}
There is a slight difference between this expression and the Lagrangian adopted in the 
Refs.~\cite{Liang:2016tqc,Ge:2017ttc}. Here we replace the domain wall tension 
$\sigma=8f_{a}^2 m_{a}$ with the effective domain wall tension $\sigma_{\rm eff}=\kappa\cdot\sigma$. 
The phenomenological parameter $\kappa$ accounts for the difference between the domain wall tension 
of a nugget $\sigma_{\rm eff}$ and that of a planar domain wall $\sigma$~\cite{Ge:2017idw}. In general, 
the effective domain wall tension $\sigma_{\rm eff}$ is smaller than $\sigma$ with $0<\kappa<1$
\footnote{There are two main reasons for the difference between $\sigma_{\rm eff}$ and $\sigma$, which 
are discussed in details in~\cite{Ge:2017idw}. We briefly summarize the two reasons here. The first 
reason is that nuggets with baryon charge accumulated inside will finally become stable in CS 
phase. Thus, in our case the axion domain wall solution interpolates between topologically distinct 
vacuum states in hadronic (outside the nugget) and CS (inside) phases, in contrast to a conventional 
axion domain wall which interpolates between distinct hadronic vacuum states. The chiral condensate 
may or may not be formed in CS phase, which could strongly make the topological susceptibility in 
the CS phase much smaller than in the conventional hadronic phase.  The second reason is 
that $\sigma=8f_{a}^2 m_{a}$ is derived using the thin-wall approximation, which could be badly 
violated in the case of the closed domain wall when the radius and the  width of the wall are at 
the same order of magnitude. This effect is expected to drastically reduce the domain wall tension.}.   
$\Delta P$ is the pressure difference inside and outside the nugget, which is~\cite{Liang:2016tqc}
\begin{equation}\label{eq:pressure}
\begin{aligned}
    \Delta P
    =&P_{\rm in}^{\rm (Fermi)}+P_{\rm in}^{\rm (bag~constant)}-P_{\rm out}\\
    =&\frac{g^{\rm in}}{6\pi^2}\int_{0}^{\infty}\frac{k^3dk}{{\rm exp}(\frac{k-\mu}{T})+1}-E_{B}
    \Theta(\mu-\mu_{1})\left(1-\frac{\mu_{1}^2}{\mu^2}\right)\\
    &-\frac{\pi^2 g^{\rm out}T^4}{90}.
\end{aligned}
\end{equation}
The first term in the right-hand side of eq.~(\ref{eq:pressure}) is the Fermi pressure inside the nugget. 
The second term is the contribution from the MIT bag model with the famous ``bag constant" 
$E_{B}\sim (150~{\rm MeV})^4$. $\Theta$ is the unit step function which implies this term turns on 
at large chemical potential $\mu>\mu_{1}$ when the nugget is in CS phase, while it vanishes at 
small chemical potential $\mu<\mu_{1}$ when the nugget is in hadronic phase. The parameter 
$\mu_{1}$ is estimated to be $\sim 330$~MeV~\cite{Zhitnitsky:2002qa} when the baryon density is 
close to the nuclear matter density. The third term is the pressure from quark gluon plasma (QGP) 
at high temperature outside the nugget. The parameter 
$g^{\rm out}\simeq (\frac{7}{8}4N_{c}N_{f}+2(N_{c}^2-1))$ is the degeneracy factor of the QGP phase. 

From the Lagrangian~(\ref{eq:lagrangian}), we obtain the equation of motion (eq. (\ref{eq:ode1}))
\begin{equation}\label{eq:ode1_appendix}
    \sigma_{\rm eff}\ddot{R}=-\frac{2\sigma_{\rm eff}}{R}-\frac{\sigma_{\rm eff}\dot{R}^2}{R}
    +\Delta P-4\eta\frac{\dot{R}}{R}-\dot{\sigma_{\rm eff}}\dot{R}.
\end{equation}
Following the same procedure of the Ref.~\cite{Liang:2016tqc}, we insert the QCD viscosity term 
$\sim4\eta\frac{\dot{R}}{R}$ to effectively describe the friction for the domain wall bubble oscillating 
in an unfriendly environment. The difference from the equation of motion in Ref.~\cite{Liang:2016tqc} 
is that here we have an extra term  $\sim\dot{\sigma_{\rm eff}}\dot{R}$. This term occurs because 
the tension $\sigma_{\rm eff}$ itself is a function of time since 
$\sigma_{\rm eff}=\kappa\cdot8f_{a}^2m_{a}(t)$. We treat the axion mass $m_{a}$ more precisely 
for it is a time (temperature)-dependent function rather than a constant. The parameter $f_{a}$ 
and the function $m_{a}(t)$ can be obtained from the axion model. Then with an appropriate 
choice of $\kappa$, we can get how $\sigma_{\rm eff}$ evolves with the cosmological time.

According to the Refs.~\cite{Liang:2016tqc,Ge:2017ttc}, the baryon charge accumulated on the 
wall is 
\begin{equation}\label{eq:Bwall}
B_{\rm wall}=g^{\rm in}\cdot 4\pi R^2\cdot\int\frac{d^2 k}{(2\pi)^2}\frac{1}{{\rm exp}(\frac{k-\mu}{T})+1},
\end{equation}
where $R$ is the radius of the nugget; $g^{\rm in}=2N_{c}N_{f}\simeq12$ and $\mu$ are 
respectively the degeneracy factor and the chemical potential of the baryon charge in the vicinity of 
the wall. The accumulated baryon charge $B_{\rm wall}$ is assumed to be a 
constant~\cite{Liang:2016tqc}, which can be expressed as 
\begin{equation}\label{eq:ode2}
\frac{d}{dt}B_{\rm wall}(t)=0.
\end{equation}
In principle, we can get the time evolution of the nugget by numerically solving the two ordinary differential equations eqs. (\ref{eq:ode1_appendix}) and (\ref{eq:ode2}). However, before we do the numerical calculations, we can get some profound analytic results from above equations. 

The nugget starts evolution at $T_{0}$ with the initial chemical potential of the baryon charge on the wall being approximately zero $\mu_{0}\simeq0$. From eq.~(\ref{eq:Bwall}), we can get the initial baryon charge accumulated on the wall
\begin{equation}\label{eq:Bwall_1}
    B_{\rm wall}(T=T_{0})\simeq \frac{\pi^2}{6}g^{\rm in}R_{0}^2 T_{0}^2.
\end{equation}
Then the nugget completes its formation at $T_{\rm form}$ when the nugget stops oscillating with $\dot{R}(t)\simeq0$, $\ddot{R}(t)\simeq0$, $\dot{\mu}(t)\simeq0$. All features of the nugget (radius, chemical potential, etc.) 
should remain almost constant after the formation point $T=T_{\rm form}$ until the very end 
$t\rightarrow \infty$ ($T\rightarrow0$). Thus, with $R_{\rm form}\simeq R(T=0)$ and $\mu_{\rm form}\simeq \mu(T=0)$ we get\begin{equation}\label{eq:Bwall_2}
\begin{aligned}
    B_{\rm wall}(T=0)&\simeq g^{\rm in}\cdot 4\pi R_{\rm form}^2\cdot\int_{0}^{\mu_{\rm form}}
    \frac{d^2 k}{(2\pi)^2}\\
    &\simeq g^{\rm in} R_{\rm form}^2\mu_{\rm form}^2.
    \end{aligned}
\end{equation}
According to (\ref{eq:ode2}), $B_{\rm wall}$ is conserved during evolution. Therefore, equating 
(\ref{eq:Bwall_1}) with (\ref{eq:Bwall_2}) we arrives at
\begin{equation}\label{eq:R_1}
\frac{R_{\rm form}^2}{R_{0}^2}=\frac{\pi^2}{6}\cdot\frac{ T_{0}^2}{\mu_{\rm form}^2}.
\end{equation}
Also, with all the derivative terms vanishing after $T_{\rm form}$, from eq.~(\ref{eq:ode1_appendix}) we get
\begin{equation}\label{eq:R_2}
R_{\rm form}\simeq R(T=0) \simeq \frac{2\sigma_{\rm eff}(T=0)}{\Delta P(T=0)},
\end{equation}
where the pressure difference $\Delta P(T=0)$ can be obtained from eq.~(\ref{eq:pressure})
\begin{equation}\label{eq:pressure0}
    \Delta P(T=0)\simeq\frac{g^{\rm in}\mu_{\rm form}^4}{24\pi^2}-E_{B}
    \left(1-\frac{\mu_{1}^2}{\mu_{\rm form}^2}\right).
\end{equation}
We notice that $R_{\rm form}$ and $\mu_{\rm form}$ are completely solvable from  eqs.~(\ref{eq:R_1}) 
to (\ref{eq:R_2}), and they are determined by the initial values $R_{0}$, $T_{0}$. 

An important feature of nugget evolution is that baryon charges not only occur on the wall of the nugget, 
but also they accumulate in the bulk of the nugget. The chemical potential $\mu$ on the wall gradually 
increases due to nugget contraction. As a consequence, the chemical potential in the bulk of the nugget 
increases keeping equilibrium with the chemical potential on the wall, which causes the accumulation 
of baryon charges in the bulk of the nugget. As explained in the Ref.~\cite{Liang:2016tqc}, the net flux 
of baryons entering and leaving the nugget $\Delta\Phi\equiv\Phi_{\rm in}-\Phi_{\rm out}$ is negligibly 
small, but the sum of these two fluxes $\left<\Phi\right>\equiv\frac{1}{2}(\Phi_{\rm in}+\Phi_{\rm out})$ 
is very large and the nugget can entirely refill its interior with fresh particles within a few oscillations.  
The high exchange rate $\left<\Phi\right>$ implies that the entire nugget could quickly reach 
chemical equilibrium. The result is that the initially baryonically neutral nugget evolves into one 
completely filled with quarks (or antiquarks for anti-nuggets). 
Therefore, we expect the nugget becomes stable at $T_{\rm form}$ with the entire nugget in 
equilibrium, with the same chemical potential $\mu_{\rm form}$. Thus, the total baryon charge 
carried by the stable nugget is 
\begin{equation}
    B\simeq g^{\rm in}\cdot \frac{4\pi}{3}R_{\rm form}^3\cdot\int_{0}^{\mu_{\rm form}}
    \frac{d^{3}k}{(2\pi)^3}\simeq\frac{2}{9\pi}g^{\rm in}R_{\rm form}^3\mu_{\rm form}^3.
\end{equation}
Then, using eq.~(\ref{eq:R_1}) we can express $B$ as
\begin{equation}\label{eq:Brelation_appendix}
    B\simeq \frac{\pi^2}{27\sqrt{6}}g^{\rm in}R_{0}^3 T_{0}^3\equiv K\cdot R_{0}^3 T_{0}^3,
\end{equation}
where $K\equiv\frac{\pi^2}{27\sqrt{6}}g^{\rm in}$ is a constant introduced for convenience. This relation is applied in section~\ref{sec:distribution} to calculate the baryon charge distribution of nuggets. 

Next, we are going to numerically solve the differential equations~(\ref{eq:ode1_appendix}) and (\ref{eq:ode2}) to get the nugget evolution. To numerically solve the two equations, we first need 
to know how the effective domain wall tension $\sigma_{\rm eff}(t)=\kappa\cdot 8 f_{a}^2 m_{a}(t)$ 
evolves as a function of time. One of the most updated results of axion mass $m_{a}(T)$ is based 
on the high-temperature lattice QCD~\cite{Borsanyi:2016ksw}. The topological susceptibility of 
QCD, $\chi(T)$, is plotted in Figure 2 in the Ref.~\cite{Borsanyi:2016ksw} as a function of the 
cosmological temperature $T$. The data points of the figure is also provided in Table 9 in the
 Supplementary Information of the same paper, by fitting which we get the expression of $\chi(T)$ as
\begin{equation}\label{eq:axion_chi}
\begin{aligned}
\frac{\chi(T)}{{\rm MeV}^4}=&3.27\times10^7\Theta[T-150{\rm MeV}]\\
&+\Theta[150{\rm MeV}-T]\frac{3.94\times10^{24}}{(T/{\rm MeV})^{7.85}},
\end{aligned}
\end{equation}
where $\Theta$ is the unit step function. Then we can get the axion mass using the relation
\begin{equation}\label{eq:axion_mass}
    m_{a}(T)=\frac{\chi^{1/2}(T)}{f_{a}}.
\end{equation}
Eqs.~(\ref{eq:axion_chi}) and (\ref{eq:axion_mass}) explicitly show that before the QCD transition 
the axion mass increase rapidly with the exponent $\beta=7.85/2=3.925$ as the cosmological 
temperature decreases (see Ref.~\cite{Wantz:2009it} for a similar result). Then the axion acquires 
its asymptotic mass near the QCD transition and remains constant after that. 

The cosmological time-temperature relationship is also useful in our numerical calculations, which 
in the radiation-dominated era is well known as 
\begin{equation}
    \frac{T(t)}{1 {\rm MeV}}\simeq1.56g_{\star}(T)^{-\frac{1}{4}}(\frac{1 {\rm sec}}{t})^{\frac{1}{2}},
\end{equation}
where $g_{\star}(T)$ is the effective degrees of freedom of all relativistic particles at temperature $T$. Since the major part of the nugget evolution is after the QCD transition, we treat $g_{\star}(T)$ as a constant for simplicity with $g_{\star}=17.25$ (see e.g.~\cite{baumann:2014cosmology}) as in the hadronic phase.

In present work, we are not going to solve the nugget evolution with the parameters such as 
$\kappa$, $f_{a}$, $T_{0}$ and $R_{0}$ taking many different values. Instead, we solve it with 
these parameters taking a group of reasonable values as an example. They are taken as 
$\kappa=0.04$, $f_{a}=10^{10}$ GeV, $T_{0}=200$ MeV and $R_{0}=6\times10^{-4}$~cm.
\exclude{
The typical radius of stable nuggets 
is about $R_{\rm form}\sim(10^{-5}-10^{-6}{\rm~cm})$, and it is usually one order smaller than the 
initial radius $R_{\rm form}\sim 0.1 R_{0}$~\cite{Liang:2016tqc,Ge:2017ttc}. Thus, as an example, 
we choose the initial radius as $R_{0}=10^{-4}$~cm. 
We emphasize that here we just choose one 
group of values of $\kappa$, $f_{a}$, $T_{0}$ and $R_{0}$ as an example to show the numerical 
result of nugget evolution. 
}
Of course the group of parameters can vary a lot, but it is not the subject 
of present work to numerically study how these parameters taking different values will affect the 
nugget evolution.

As we discussed in section~\ref{sec:RTE}, when we try to solve the two differential equations, 
we immediately meet the multiple-scale problem $\omega\tau\sim10^{10}$ which implies that 
the nugget will not settle down  until after billions of oscillations. The enormous number of oscillations 
make the code extremely time-consuming, which makes it almost impossible to completely solve the 
system numerically. To make the numerical solution feasible, the QCD viscosity term $\eta$ was artificially 
enlarged 8 or 9 orders in our previous calculations in Refs.~\cite{Liang:2016tqc,Ge:2017ttc}. 

In this 
paper, we  adopt a different numerical method coined as  ``envelope-following method" to solve the 
system, with the viscosity term keeping its physical magnitude $\eta\sim \Lambda_{\rm QCD}^3$ when parameter $\omega\tau\sim 10^{10}$ assumes its very large physical value. We believe this will make our 
numerical result of nugget evolution more trustworthy. The motivation of using the envelope-following 
method and how it works are briefly explained as follows.

We notice that although a nugget is oscillating very fast during evolution, the amplitude of oscillation 
is decreasing very slowly for each given cycle. The peaks of oscillations in fact form a 
``smooth'' line which we call \textit{envelope}. We realize that if we can find a way to numerically solve 
the envelope, then it is unnecessary to solve the full details of all oscillations. The envelope-following 
method turns out to be very  beneficial for our problem  of study of  the nugget oscillations. The method 
is efficient in solving highly oscillatory ordinary differential equations, which is illustrated in 
Ref.~\cite{petzold1981efficient}. We briefly summarize the basic idea here.

We start with the initial conditions $R=R_{0}, \dot{R}=0$ (also $\mu=\mu_{0}$), which corresponds to 
the first peak of $R$ oscillations and we label this peak as point $a$. Then we solve the differential 
equations until we get the next peak $b$ of $R$ oscillations, which should be slightly smaller than the 
first peak. This step is very fast since we solve the equations just for one oscillation. Joining points $a$ 
and $b$, we get a secant line. This secant line then is used to project the solution to point $a'$ which 
is many oscillations away. Starting with $a'$ as the new peak, we solve the differential equations until 
we get the next peak $b'$... We repeat the above procedure of drawing the secant line, projecting 
the solution and finding the next peak. After several projections, we get the upper envelope of 
$R$ oscillations. Using the same method, we can find the lower envelope of $R$ oscillations and also 
the envelopes of $\mu$ oscillations. We should point out that, although the details of oscillations are 
not important for us, we can recover them locally if we substitute the corresponding envelope 
information into the differential equations as the initial conditions.

We plot the numerical result of nugget evolution solved by envelope-following method in 
Fig.~\ref{fig:RTE} with the group of parameters chosen above.  In Fig.~\ref{fig:RTE}, we see that in 
this case the nugget complete its evolution at $\sim 40$ MeV. The formation chemical potential is $\sim 450$ MeV, well above the threshold of CS phase 
$\mu_{1}=330$ MeV. Also, we see that the theoretical analysis of $R_{\rm form}$ and $\mu_{\rm form}$ (denoted 
as dashed blue line and dashed orange line respectively) from eqs.~(\ref{eq:R_1}) and (\ref{eq:R_2}) 
matches the numerical result pretty well, which verifies the validity of the two equations and further the 
relation~(\ref{eq:Brelation_appendix}) $B\propto R_{0}^3 T_{0}^3$.

\section{Calculations of $N(B)$ and $dN/dB$}\label{appen:NB}
\begin{figure}
    \centering
    \includegraphics[width=\linewidth]{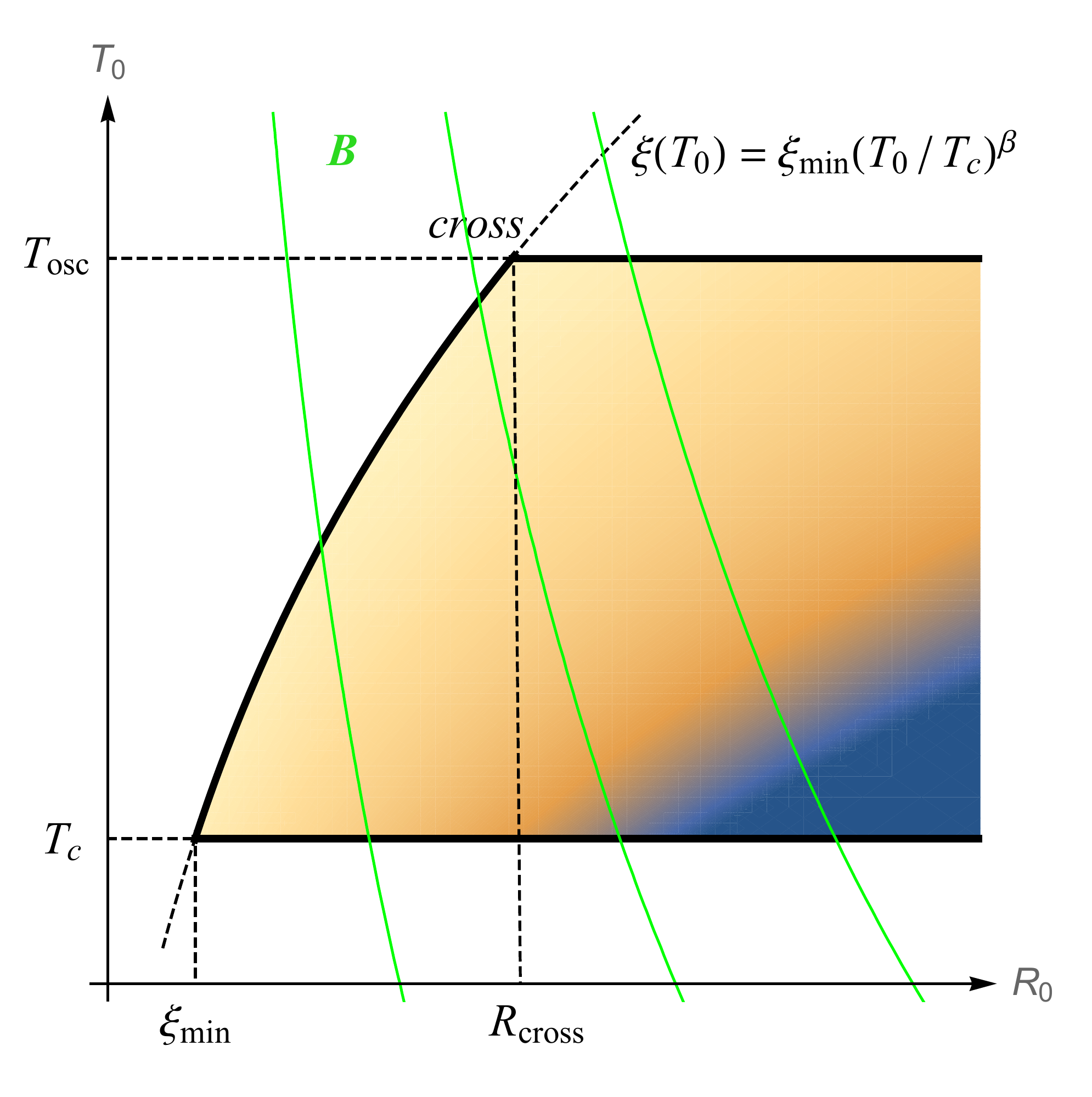}
    \caption{Parameter space of $R_{0}$ and $T_{0}$. The coloured region represents the allowed 
    ($R_{0}, T_{0}$) for the formation of closed domain walls initially. Different colours represent different 
    magnitudes of $f(R_{0}, T_{0})$ which decreases from the light yellow part to the deep blue part 
    (gradually away from the correlation length $\xi(T_{0})$). The green lines are the contour lines of 
    $B$, i.e. each line corresponds to the same value of $B$, with $B$ increasing from the left lines to 
    the right.}
    \label{fig:f_diagram}
\end{figure}

In this section, we will show the details of calculating eq.~(\ref{eq:NB}) and eq.~(\ref{eq:f}), to support 
the results in section~\ref{subsec:plotsNB}. We first rewrite eq.~(\ref{eq:NB}) as
\begin{equation}\label{eq:NB1}
   N(B)=N_{0}P\int_{T_{c}}^{T_{c}
   \cdot\left(\frac{B}{K\xi_{\rm min}^3T_{c}^3}\right)^{\frac{1}{3(\beta+1)}} }
   dT_{0}\int_{\xi(T_{0})}^{\left(\frac{B}{KT_{0}^3}\right)^{\frac{1}{3}}}dR_{0}~f
\end{equation}
with the limits of integration written explicitly, which can be explained as follows. The 
integral~(\ref{eq:NB1}) is performed over the region $KR_{0}^3T_{0}^3\leq B$ with the constraints 
$T_{c}\lesssim T_{0}\lesssim T_{\rm osc}$ and $R_{0}\gtrsim\xi(T_{0})$ from the model of $T_{0}$ and 
$R_{0}$ distributions. We show the region of integration in Fig.~\ref{fig:f_diagram}, where the parameter 
space $T_{c}\lesssim T_{0}\lesssim T_{\rm osc}$ and $R_{0}\gtrsim\xi(T_{0})$ is represented by the 
coloured region. The green lines are the contour lines of $B$ with $KR_{0}^3T_{0}^3=B$ for different 
values of $B$. Then the region of integration is the area enclosed by the solid black lines and one of the 
green lines (to the left of the green line), from which we can obtain the limits of integration. The lower 
limit of $R_{0}$ is $R_{\rm lower}=\xi(T_{0})$; the upper limit of $R_{0}$ is on the green line 
$R_{\rm upper}=[B/(KT_{0}^3)]^{1/3}$; the lower limit of $T_{0}$ is $T_{c}$. The upper limit of $T_{0}$ 
is a little complicated: It could either be the intersection of the line $\xi(T_{0})$ and the green line which 
is $T_{\rm upper}=T_{c}\cdot[B/(K\xi_{\rm min}^3T_{c}^3)]^{1/3(\beta+1)}$, or simply 
$T_{\rm upper}=T_{\rm osc}$, depending on different values of $B$. However, we are not likely to have 
the chance to use the latter case where $T_{\rm upper}=T_{\rm osc}$ as the upper limit of $T_{0}$, which is 
explained as follows.   

\begin{figure}
\subfloat[]{\includegraphics[width=\linewidth]{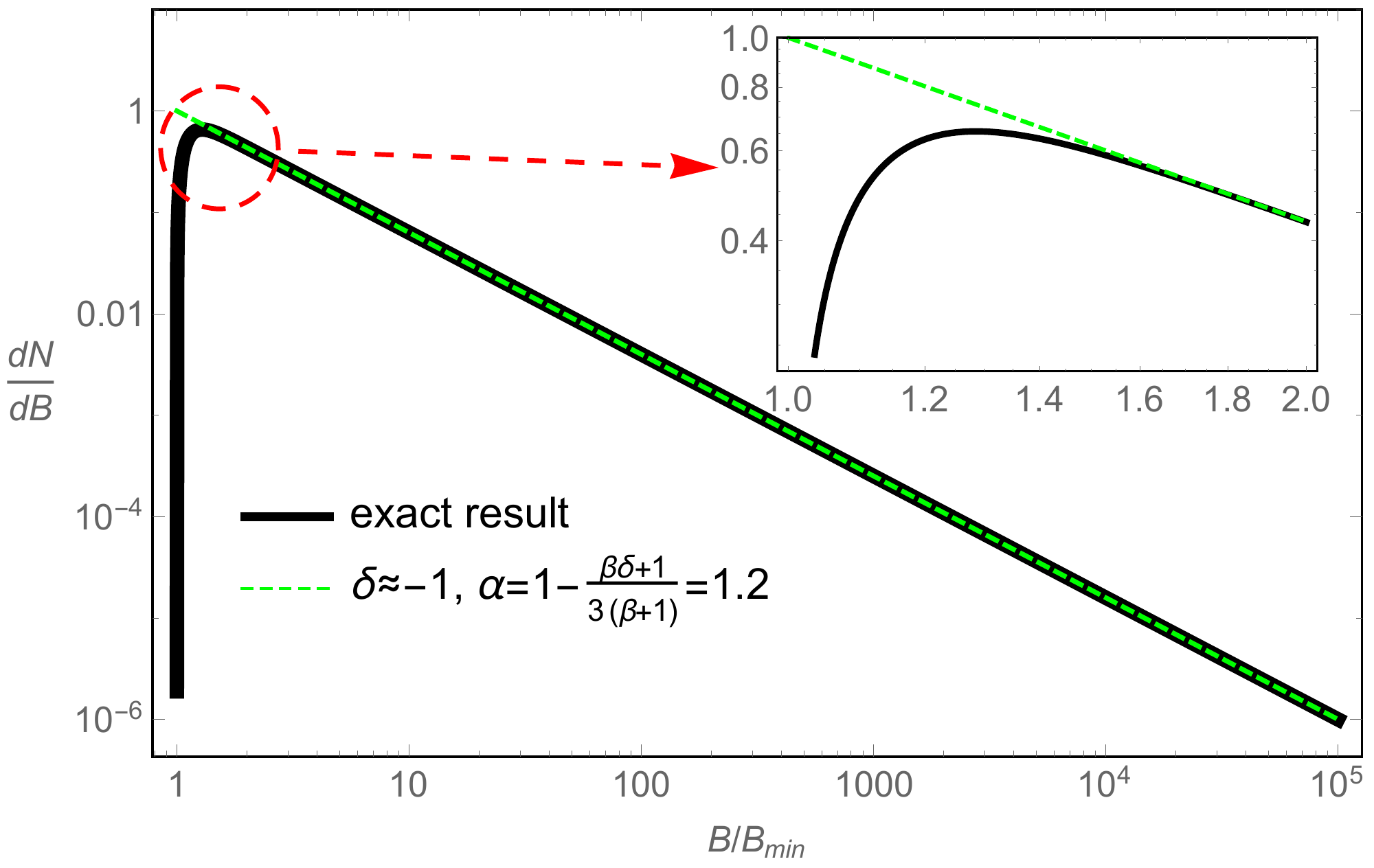}\label{fig:alpha1}}
\newline
\subfloat[]{\includegraphics[width=\linewidth]{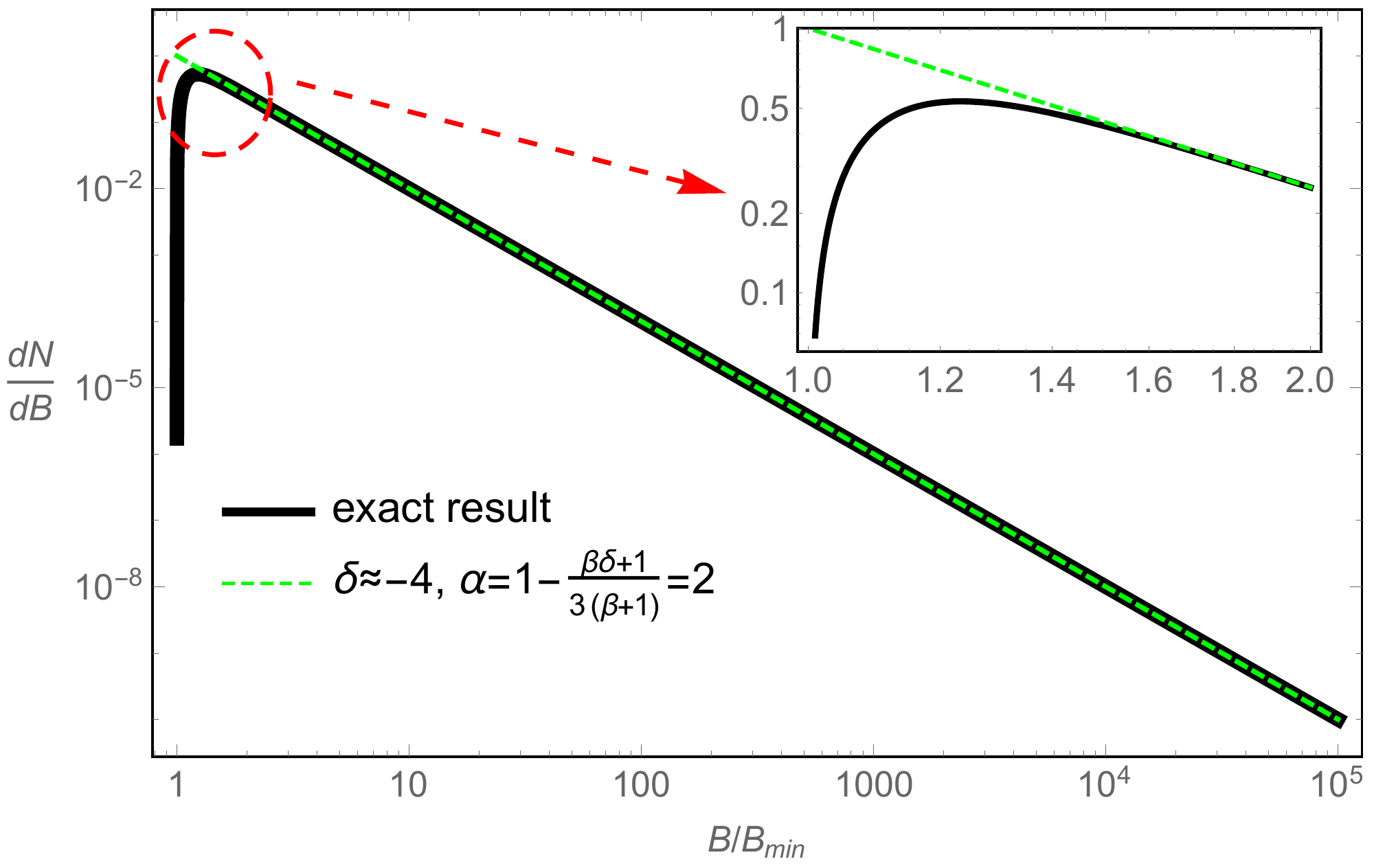}\label{fig:alpha2}}
\caption{The relation between $dN/dB$ and $b\equiv B/B_{\rm min}$. We choose $\tau=2$, 
$\lambda=10$ and $\beta=3.925$ for both subfigures. The difference between (a) and (b) is the values 
of $\delta$. In (a), $\delta\approx-1$ and thus $\alpha =1.2$; in (b), we choose $\delta\approx-4$ 
to make $\alpha =2$. The solid black line and dashed green line in each subfigure 
represents eq.~(\ref{eq:f1A}) and eq.~(\ref{eq:slope_appendix}) respectively (the prefactor 
$N_{0}P/B_{\rm min}$ in the two equations is rescaled to make $dN/dB$ completely shown in the range 
from 0 to 1 for illustrative purpose).} 
\label{fig:alpha}
\end{figure}

If we want the upper limit of $T_{0}$ in the integration to be $T_{\rm osc}$, then $B$ has to be larger 
than $B_{\rm cross}=K\xi^3(T_{\rm osc})T_{\rm osc}^3$, the value of $B$ at the crossing point where the line 
$\xi(T_{0})$ intersects the horizontal line $T_{0}=T_{\rm osc}$. We should compare $B_{\rm cross}$ with 
the minimal baryon charge $B_{\rm min}=K\xi_{\rm min}^3T_{c}^3$ which corresponds to the closed 
domain wall initially forming at $T_{0}=T_{c}$ with $R_{0}=\xi_{\rm min}$. We get 
\begin{equation}
B_{\rm cross}=\left(\frac{T_{\rm osc}}{T_{c}}\right)^{3(\beta+1)}B_{\rm min}\simeq 10^{15}B_{\rm min}
\end{equation}
where we approximate it using $T_{\rm osc}/T_{c}\simeq10$ and $\beta\simeq3.925$. We see that the range is $15$ orders of magnitude wide, which is large enough for us to match the baryon charge distribution of nuggets with the energy distribution of solar nanoflares. Therefore, we choose $T_{0}$ is $T_{\rm upper}=T_{c}\cdot[B/(K\xi_{\rm min}^3T_{c}^3)]^{1/3(\beta+1)}$ as the upper limit in  eq.~(\ref{eq:NB1}).
\exclude{
However, from 
section~\ref{subsec:plotsNB} we know that the baryon charge distribution $dN/dB$ starts to follow a 
power-law from $B\sim1.3B_{\rm min}$. The range $1.3B_{\rm min}\lesssim B\lesssim 10^{8}B_{\rm min}$ 
is large enough to cover the power-law distribution of the energy of solar nanoflares (whose energy 
window is about six orders wide, see section \ref{subsec:plotsNB}), which is far below 
$B_{\rm cross}\simeq 10^{15}B_{\rm min}$. Thus, in the range 
$10B_{\rm min}\lesssim B\lesssim 10^{8}B_{\rm min}$ that we are interested in, the upper limit of 
$T_{0}$ is $T_{\rm upper}=T_{c}\cdot[B/(K\xi_{\rm min}^3T_{c}^3)]^{1/3(\beta+1)}$ as we mentioned 
above, which is also the one chosen in eq.~(\ref{eq:NB1}). 
}

Next, we are going to calculate eq.~(\ref{eq:NB1}). Using the definitions $r=R_{0}/\xi_{\rm min}$ and 
$u=T_{0}/T_{c}$, we rewrite eq.~(\ref{eq:f}) in a more concise way
\begin{equation}
f(r,u)=\frac{1}{\xi_{\rm min}T_{c}}\cdot u^{3\beta(\tau-1)+\beta\delta} \cdot r^{2-3\tau}
\cdot {\rm e}^{-\lambda r^2 u^{-2\beta}}.
\end{equation}
Substituting $f(r,u)$ into eq.~(\ref{eq:NB1}) and using the definition $b=B/B_{\rm min}$, we arrive at
\begin{equation}
\begin{aligned}
    N(b)&=N_{0}P\cdot T_{c}\xi_{\rm min}\int_{1}^{b^{\frac{1}{3(\beta+1)}} }du 
    \int_{u^{\beta}}^{u^{-1}b^{\frac{1}{3}}}dr~f(r,u)\\
    &=N_{0}P\int_{1}^{b^{\frac{1}{3(\beta+1)}} }du 
    \int_{u^{\beta}}^{u^{-1}b^{\frac{1}{3}}}dr~\left[u^{3\beta(\tau-1)+\beta\delta}\right.\\
    &~~~~\left.\times r^{2-3\tau}\cdot {\rm e}^{-\lambda r^2 u^{-2\beta}}\right],
\end{aligned}
\end{equation}
from which we further get
\begin{equation}\label{eq:f1A}
    \begin{aligned}
    \frac{dN}{dB}&=\frac{1}{B_{\rm min}}\frac{dN}{db}=\frac{N_{0}P}{3B_{\rm min}}\cdot b^{-\tau}\\
    &~~~~\times\int_{1}^{b^{\frac{1}{3(\beta+1)}} }du~u^{3(\beta+1)(\tau-1)+\beta\delta}
    \cdot{\rm e}^{-\lambda b^{\frac{2}{3}} u^{-2(\beta+1)}}.
    \end{aligned}
\end{equation}
This can be further simplified using
\begin{equation}\label{eq:fsimp}
\begin{aligned}
    &~~~~\int_{1}^{b^{\frac{1}{3(\beta+1)}}}dt~u^m{\rm e}^{-\lambda b^{2/3} u^{-n}}\\
    &=\frac{1}{n}(\lambda b^{2/3})^{\frac{1+m}{n}}\cdot\left.
    \Gamma(-\frac{1+m}{n},\lambda b^{\frac{2}{3}}u^{-n})\right|_{u=1}^{u=b^{\frac{1}{3(\beta+1)}}}\\
    &\simeq \frac{1}{n}(\lambda b^{2/3})^{\frac{1+m}{n}}\cdot\Gamma(-\frac{1+m}{n},\lambda),~{\rm for}
    ~b\gg1,
\end{aligned}
\end{equation}
where $m\equiv 3(\beta+1)(\tau-1)+\beta\delta$ and $n\equiv 2(\beta+1)$; 
$\Gamma(s,x)=\int_{x}^{\infty}t^{s-1}{\rm e}^{-t} dt$ is the \textit{incomplete gamma function}. To obtain 
the last approximate equality, we neglect the term $\Gamma(-\frac{1+m}{n},\lambda b^{2/3})$ since it 
is far smaller than the term $\Gamma(-\frac{1+m}{n},\lambda)$ for $b\gg 1$. The condition $b\gg1$ 
is satisfied in a wide range of $B$ which is generally several orders larger than $B_{\rm min}$.  
Substituting (\ref{eq:fsimp}) into (\ref{eq:f1A}), we arrive at 
\begin{equation}\label{eq:slope}
    \frac{dN}{dB}=\frac{N_{0}P}{3B_{\rm min}}\frac{1}{n}\lambda^{\frac{1+m}{n}}
    \Gamma(-\frac{1+m}{n},\lambda)\cdot b^{-1+\frac{\beta\delta+1}{3(\beta+1)}},~~b\gg1.
\end{equation}
We see that $dN/dB$ follows a power-law distribution
\begin{equation}\label{eq:slope_appendix}
\frac{dN}{dB}\propto b^{-\alpha},~{\rm with}~\alpha =1-\frac{\beta\delta+1}{3(\beta+1)},~b\gg1
\end{equation}
which verifies the relation~(\ref{eq:slope_prop}). 
The finite-cluster parameters $\tau$ (contained in $m$) and $\lambda$ that we have discussed in 
section~\ref{subsec:R0} only affect the relative magnitude of $dN/dB$, but not the slope of the power-law 
distribution $-\alpha $. 

The parameter $\beta$ describing the relation between axion 
mass and cosmological temperature is well calculated~\cite{Borsanyi:2016ksw} (see also 
Appendix~\ref{appen:evolution} for more details). The other parameter $\delta$ from the model of 
$T_{0}$ distribution eq.~(\ref{eq:T0distribution}) is relatively adjustable, which can result in different 
slopes of the power-law distribution $dN/dB$. This parameter which may have any sign, plus or minus, describes 
the distribution of the bubble formation.  As we explained in the text, the positive  sign of $\delta$ corresponds to the 
preference of the bubble formation close to $T_{\rm osc}$, while the negative $\delta$ corresponds to a preferred bubble formation close to $T_c$ with much stronger tilt of the axion potential.

We plot the baryon charge distribution of nuggets in Fig.~\ref{fig:alpha}. We choose $\tau=2$, 
$\lambda=10$ and $\beta=3.925$ for both subfigures. The difference between them is the value of 
$\delta$ which is highly underdetermined.  In Fig.~\ref{fig:alpha1}, we choose
$\delta\approx -1$ to make $\alpha =1.2$. The solid black line is the plot of eq.~(\ref{eq:f1A}), which represents the exact result of the 
distribution. As a comparison, we also plot the approximate relation~(\ref{eq:slope_appendix}) represented 
by the dashed green line, which is straight in the $\log$-$\log$ scale. We see that the approximate relation~(\ref{eq:slope_appendix})  matches the exact result eq.~(\ref{eq:f1A}) pretty well after the turning point where the condition $b\gg1$ becomes valid. 
\exclude{
In this case $\delta=0$, the exponent 
of the green line is $\alpha_{\rm nugget}=0.93$ according to eq.~(\ref{eq:alpha_nugget}). We also see 
two vertical dashed black lines in each subfigure. The first line represents $b=1$, i.e. $B=B_{\rm min}$ 
which corresponding to the starting point of the solid black line eq.~(\ref{eq:f1A_copy}). The second 
line locates at the turning point of the distribution, after which the the approximate 
relation~(\ref{eq:slope_prop}) matches pretty well with the exact result eq.~(\ref{eq:f1A_copy}). Recall 
that we obtain the power-law (\ref{eq:slope_prop}) under the condition $b\gg1$. From the figure we see 
that this condition is easily satisfied even for $b\simeq1.3$ (or $1.23$ in the second subfigure) which is 
just slightly larger than $1$.
}
In Fig.~\ref{fig:alpha2}, we consider the case $\delta\approx-4$ which 
corresponds to $\alpha=2$. All other ingredients are the same as the first subfigure.

\section{On the estimate of parameter $\kappa(T)$ during pre-BBN evolution}\label{kappa}
The main goal of this Appendix is to make an order of magnitude estimate of the parameter $\kappa(T)$ 
which is an important element of our analysis  in Sect. \ref{ssec:PreBBN}. In simple quantum mechanics terms the parameter 
$\kappa(T)$ is defined as the transmission coefficient for the baryon to enter and annihilate inside the nugget, while the reflection coefficient $[1-\kappa(T)]$  describes the probability for the baryon's reflection off the sharp surface of the nugget.
One should emphasize that the system cannot be formulated in the simple terms normally used for a one particle QM description.   
Instead, a proper study  of this phenomenon would require a non-perturbative QFT description as  many body effects play a crucial role in the computations. This is because the key  element of the analysis must be a description of  physics at the interface between hadronic phase (as the proton's wave function is formulated  in terms of the quarks) and a CS phase characterized by diquark vacuum  condensate. One should emphasize that the diquark condensate    is not a local object, but a complicated coherent superposition of quarks similar to Cooper pair in conventional superconductors.  

In spite of the complicated  annihilation pattern of a baryon  in the hadronic phase with diquarks in the CS phase 
  as highlighted above  one can   easily carry out  a simple,  order of magnitude, estimate   
  demonstrating  that $\kappa(T)\ll 1$, which is precisely the main goal of this Appendix. To simplify things we separate the 
  suppression factor $\kappa$ in two pieces: $\kappa\sim \kappa_1\cdot \kappa_2$, where $\kappa_1$ is defined as the dynamical suppression factor, while $\kappa_2$ is defined as the kinematical  suppression factor, see below. 
  
  The dynamical  suppression factor $\kappa_1$ can be easily understood from the internal structure of the axion domain wall represented by heavy $\eta'$ field which   accompanies the axion field in the AQN construction. The corresponding  very sharp   QCD structure has a width $\sim m^{-1}_{\eta'}\ll \Lambda_{\rm QCD}^{-1}$, see \cite{Forbes:2000et}. Therefore, one should expect some  suppression due to  a sharp potential 
  \be
  \label{sharp}
   \kappa_1(T)\sim \left(\frac{E}{\Lambda_{\rm QCD}}\right)^3\sim\left(\frac{T}{\Lambda_{\rm QCD}}\right)^3, 
\ee
where we assume that a typical energy $E$ of the incoming three quarks making the  proton   is order of $T$, while typical strength of the ``potential" is order of $\Lambda_{\rm QCD}$. We emphasize that we should formulate the problem in terms of quarks (rather than in terms of a single   proton's wave function)  because the annihilation pattern should include 3 antiquarks from a different CS phase. The suppression parameter $\kappa_1(T)$ represents the  probability for simultaneous transmission of 3 quarks. 
We note that the factor $E/\Lambda_{\rm QCD}\ll 1$ in  (\ref{sharp}) for a sharp surface   is a very generic QM feature, which is not  even  sensitive to the sign of interaction. It can    be  
checked with a simple QM problem of scattering of a low energetic particle with $E\rightarrow 0$ on a $\delta (x)$ potential.   

There is another suppression factor $\kappa_2(T)$  related to  a strong mismatch between  wave functions of quarks in hadronic phase  and antiquarks in CS phase.  The corresponding suppression  always has  a  factor $1/N!$ for $N$ constituents which is $N=3$ for proton\footnote{We put generic factor $N$  instead of $3$ in eq. (\ref{permutations}) on purpose to emphasize that the annihilation process of $N$ different constituents must find their counterparts with precisely matching wave functions for successful annihilation.}. The $\kappa_2(T)$  also depends on  overlapping features   of the wave functions from drastically different phases (hadronic vs CS phase). Therefore, we can represent $\kappa_2(T)$ as follows
\be
\label{permutations}
\kappa_2(T)\sim \frac{1}{N!}\cdot ({\rm overlapping ~integrals}).
\ee  
Collecting estimates (\ref{sharp}) and (\ref{permutations}) together one should expect that
\be
\label{final}
\kappa(T)\lesssim 10^{-3} ~{\rm for }~ T\simeq 40~ {\rm MeV},     \Lambda_{\rm QCD}\simeq 170~ {\rm MeV}, ~~~~
\ee
which represents our final estimate for suppression factor $\kappa(T)$ used in Sect. \ref{ssec:PreBBN}.
The order of magnitude estimate (\ref{final}) is sufficient for qualitative arguments suggesting that the nuggets easily survive 
the dense and hot environment after the formation. 

It is interesting to note that the  solitons and anti-solitons  (which can be represented as coherent superposition of large $N\rightarrow \infty$ number of constituents) in condensed matter physics do not normally annihilate, but rather experience an elastic scattering, which can be thought as the  manifestation of the factor $1/N!$ in (\ref{permutations}). It is also known 
 that the anti-proton does not easily annihilate with a large nuclei (considered to be in a nuclear matter phase), but could have a life time as long as $\sim 20~ {\rm fm}/c$ instead of conventional $\sim   {\rm fm}/c$ scale  \cite{Mishustin:2004xa}.
 This suppression of annihilation can be attributed (analogous) to our ``overlapping" factor in (\ref{permutations}).

\section{The AQNs in corona: the turbulence and  effective cross section with plasma  }\label{turbulence}

The main goal of this Appendix is to explain the crucial difference between our analysis 
presented in sections \ref{sec:survival}, \ref{ssec:PreBBN} and \ref{low-T} where we argued that very few annihilation events may occur 
in early Universe. 
 At the same time, it has been argued in   \cite{Zhitnitsky:2017rop, Zhitnitsky:2018mav, Raza:2018gpb} that the nuggets of all sizes will experience  the complete annihilation  in the solar corona when the AQNs enter the solar atmosphere. There is no contradiction between these two claims.

Indeed, our estimates (\ref{eq:hot_col}) and (\ref{eq:cool_col}) indicate  that    $\Delta B\ll B$ during the entire evolution of the Universe from soon after the nugget's formation at $T\approx 40 $ MeV until the present time.   We considered two different regimes:  during the hottest period of evolution with $T_*<T<40 $ MeV then the nuggets' electrosphere is made of positrons and the annihilation of a baryon charge is highly unlikely and when the temperature drops to below $T_*\approx 20 $ keV and the electrosphere contains a significant number of protons and baryon charge annihilation may occur as estimated in section \ref{low-T}. 

This number of collisions in plasma at $T_*\approx 20 $ keV should be compared with number of collisions when an AQN enters the solar corona. The corresponding estimate goes as follows, see  \cite{Zhitnitsky:2017rop, Zhitnitsky:2018mav, Raza:2018gpb} for the details. First, we estimate the ionized charge of the nugget in terms of the internal temperature $T_I$ similar to our estimate (\ref{Q}) \exclude{with the only difference that the internal temperature $T_I$ is not the same as external temperature $T_P$ of the surrounding plasma $T_P$ }
 \be
  \label{Q1}
Q\simeq 4\pi R^2 \int^{\infty}_{z_0}  n(z)dz\sim \frac{4\pi R^2}{2\pi\alpha}\cdot \left(T_I\sqrt{2 m_e T_I}\right). ~~~~~
  \ee
Where $T_P$ is the temperature of the surrounding plasma. 
The difference between $T_P$ and $T_I$ is enormous as emphasized in \cite{Raza:2018gpb} and related to the fact the nugget propagates in the solar corona
with velocity  $v  \gtrsim 600 $ km/s (escape velocity for the Sun) which greatly exceeds the speed of sound $c_s$ in corona, i.e. the Mach number $M\equiv v/c_s\sim 10$.  It is well known that a moving body with such a large Mach number will inevitably generate the shock waves.   It is also known that a shock wave generates a discontinuity in temperature, which for large Mach numbers $M\gg 1$ can be approximated as follows  
  \be
 \label{shock1}
  \frac{T_2}{T_1}\simeq M^2\cdot \frac{2\gamma(\gamma-1) }{(\gamma+1)^2}, ~~~ \gamma\simeq 5/3, 
 \ee
where the  temperature $T_1\simeq T_P$  is identified with the temperature of the surrounding unperturbed plasma, while the high temperature $T_2$  can be thought of as the internal temperature of the nuggets $T_I$. Precisely this very high internal temperature $T_I/T_P\sim M^2$ being developed due to the shock wave makes a huge difference in comparison with the analysis in section \ref{low-T} for   the   plasma in the thermal equilibrium at $T\simeq T_*$.

The effective cross section of AQN   with surrounding protons can be estimated as $\pi R_{\rm eff}^2$, where $R_{\rm eff}$ corresponds to the distances where
protons from plasma with energy $\sim T_P$ can be captured by the nugget
 \be
 \label{momentum}
T_P \sim \frac{Q\alpha}{R_{\rm eff}}.
 \ee
 Combining (\ref{Q1}) with (\ref{momentum}) one can estimate the enhancement of the effective cross section in comparison with naive estimate $\pi R^2$ as follows
   \be
 \label{T}
  \left(\frac{R_{\rm eff}}{R}\right)^2\simeq \frac{8 (m_e T_P) R^2}{\pi}\left(\frac{T_I}{T_P}\right)^3 \sim M^6. 
 \ee
  This enhancement factor,  of course, is different for different nugget's velocities as Mach number $M$ varies with $v$. This enhancement factor  obviously also changes  with time and solar altitude as a result of a motion with the friction and  radiation. One should also emphasize that there will be  very efficient energy and momentum exchange and  heat transfer to the surrounding  plasma   as a result of nonequilibrium dynamics in the form of the  developed turbulence in vicinity of the shock wave front. It should be    contrasted with analysis in Section  \ref{low-T} when the system is in perfect thermal equilibrium.    The numerical simulations performed in \cite{Raza:2018gpb} suggest that most of the nuggets loose their entire baryon charges due to the annihilation $\Delta B\simeq B$  in vicinity of the transition region at the altitude $\sim 2000$ km where it is known that  the drastic changes in  temperature and pressure do occur. 
  
  From AQN perspective such unusual features of the  transition regions as well as the ``solar corona heating puzzle" are understood in terms of  the dark matter nuggets which  continuously hit  the solar atmosphere with the flux  which has correct magnitude    to saturate the EUV and soft x-ray radiation from the corona.

\bibliography{MDrefs-revised} 
\end{document}